\documentclass[longauth]{aa}  
\usepackage{graphicx}
\usepackage{txfonts}
\usepackage{lipsum}
\usepackage{subcaption}         
\usepackage{lscape}            
\usepackage{placeins}     
\usepackage{longtable}        
\usepackage{hyperref}
\usepackage{upgreek}   

\begin{document} 

   \title{Detection and characterisation of binary asteroid candidates through stellar occultations}
    \subtitle{}

\author{R. Lallemand\inst{1}\fnmsep\thanks{Corresponding author: raphael.lallemand@obspm.fr} \and 
J. Desmars\inst{1}\fnmsep\inst{2} \and 
B. Sicardy\inst{1} \and 
Z. Liu\inst{1} \and 
P. Tanga\inst{3} \and 
L. Liberato\inst{3} \and 
B. Carry\inst{3} \and 
A. Leroy\inst{4} \and 
Y. Kilic\inst{5}\fnmsep\inst{6} \and 
M. Assafin\inst{7}\fnmsep\inst{8}
% ...
}

\institute{ Laboratoire Temps Espace (LTE), Paris Observatory, PSL Research University, CNRS, Sorbonne University, UPMC Univ Paris 06, Univ. Lille, 77, Av. Denfert-Rochereau, 75014 Paris, France \and Institut Polytechnique des Sciences Avanc\'ees IPSA, 63b Bd. de Brandebourg, 94200 Ivry-sur-Seine, France. \and Universit\'e  C\^ote d’Azur,  Observatoire  de  la  C\^ote  d’Azur,  CNRS,Laboratoire Lagrange, Bd de l’Observatoire, CS 34229, 06304 Nice Cedex 4, France. \and Uranoscope de l'Ile de France, All. Camille Flammarion, 77220 Gretz-Armainvilliers, France \and Instituto de Astrof\'isica de Andaluc\'ia (IAA-CSIC), Glorieta de la Astronom\'ia s/n, 18008-Granada, Spain \and LIRA, Observatoire de Paris, Universit\'e PSL, Sorbonne Universit\'e, Universit\'e Paris Cit\'e, CY Cergy Paris Universit\'e, CNRS, 92190 Meudon, France \and Universidade Federal do Rio de Janeiro - Observat\'orio do Valongo, Ladeira do Pedro Ant\^onio 43, Rio de Janeiro, 20.080-090, RJ, Brazil \and Laborat\'orio Interinstitucional de e-Astronomia (LIneA), Rua General Jos\'e Cristino 77, Rio de Janeiro, 20.921-400, RJ, Brazil
% ...
}

\abstract
   {Binary asteroids provide key access to fundamental parameters of Solar System remnants and planetary formations. However, the current knowledge of binary asteroids remains strongly biased by observational limitations, and main belt binary systems are still poorly characterised since current techniques preferentially detect either widely separated binaries through direct imaging or close and bright systems via photometry and radar for near-Earth asteroids. In this context, the high-precision astrometry of the Gaia mission has revealed a new population of candidate binaries exhibiting dynamical signatures consistent with unresolved companions. Stellar occultations have therefore emerged as one of the most effective methods to confirm the binary nature of a candidate and improve the current census of intermediate-size systems.}
   {This work is part of the GaiaMoons program, and our aim with it was to characterise a sample of 357 potential binary asteroid targets and confirm or refute their binary nature. The properties of these candidates were derived from the high-precision photometric and astrometric observations provided by the Gaia satellite.} 
   {We adopted stellar occultation as the observational method to study these targets. Between October 2023 and February 2026, we successfully carried out 165 observations for 101 targets. We subsequently analysed these events in the context of the available literature and previously reported observations.}
   {Out of the 165 observations, 76 led at least to one positive observation. Among these, 33 had at least two positives for 24 objects that have undergone unprecedented occultation observation campaigns, with four objects showing indications of binary or contact binary features, namely (1127) Mimi, (35420) 1998 AG$_6$, (206) Hersilia, and (36882) 2000 SW$_{155}$. For the vast majority of these objects, the resulting dataset from all reduced observations provides unique physical and astrometric constraints, as they had never been observed through stellar occultations before. In addition, 89 observations with only negatives allowed the near environment of the targets to be probed.}
   {GaiaMoons illustrates how stellar occultation campaigns associated with Gaia observations generate a self-improving cycle to find new binary, thereby probing size and shape to constrain future observations. By standardising this approach, we deliver critical data in unexplored parameter spaces, resolving long-standing observational ambiguities.}
   
\keywords{Asteroids -- Astrometry -- Shape -- Structure -- Satellite -- Occultations.}

\maketitle
\nolinenumbers
%
\section{Introduction}
Binary and multiple systems are now recognised as a common outcome of small-body evolution in the Solar System. \cite{2006Icar..181...63P} and \cite{2002Sci...296.1445M} estimated a a fraction of small binaries among near-Earth asteroids (NEAs) at $15 \pm 4$\% . Similarly, \cite{2010Icar..207..732C} estimated a comparable binary fraction among small-diameter ($D < 10$~km) main belt asteroids (MBAs), and \cite{2008ssbn.book..345N} did so for trans-Neptunian objects (TNOs). Binary asteroids represent a key population for understanding the formation and evolution of small bodies in the Solar System. The presence of a satellite provides a natural way to determine the primary’s mass through dynamical studies of the mutual orbit, which in turn allows estimation of its bulk density and internal structure \citep{2025AJ....170..353F}. Comparisons between densities and compositions offer crucial constraints on accretion processes, collisional history, and possible differentiation \citep{2019A&A...623A.132C,2021A&A...650A.129C}. Hence, the growing interest in binary asteroids is motivated by their strong diagnostic power for constraining internal structure, density, formation mechanisms, and long-term dynamical evolution \citep{2006AREPS..34...47R,2015aste.book..355M,2025A&A...701A..42M}. 
Despite significant progress, the study of binary asteroids remains affected by strong observational biases and intrinsic limitations. Radar observations are restricted to a small number of close-approach NEAs \citep{2002aste.book..151O}, adaptive optics is limited to the largest and closest systems \citep{2021A&A...654A..56V}, and light curve inversion techniques are mainly sensitive to tightly bound binaries with favourable geometries \citep{2002aste.book..289M}. As a result, current samples are incomplete and strongly biased. Among the known binaries, there is a bimodal distribution in sizes towards small primaries smaller than 20~km and larger than 80~km, with a gap in between.\footnote{\url{https://johnstonsarchive.net/astro/asteroidmoons.html}}
\\ 

A number of large-scale studies have searched for undiscovered binary systems among known Solar System objects \citep{2007Icar..190..250P,2017A&A...608A..19K}. GaiaMoons\footnote{\url{https://www.oca.eu/fr/gaiamoons}} \citep{2024sf2a.conf..483L,2025sf2a.conf..177L,2024A&A...688A..50L} is one of these projects, and the present work contributes to this program. 
The ESA mission Gaia \citep{2016A&A...595A...1G} provides highly precise astrometric data, especially the Gaia Focus Product Release (FPR; \citep{2023A&A...680A..37G}) for Solar System objects. The combination of the FPR astrometry and the improved processing of Gaia data have opened new perspectives for the detection of binary asteroids through the systematic search for variations in the astrometric post-fit residuals due to the periodic displacement between the photocentre and the barycentre of the target (wobble), which can be induced by the gravitational perturbation from an unresolved companion \citep{2023A&A...674A..12T,2024A&ARv..32....6S,2024A&A...688L..23L}.
In particular, \cite{2024A&A...688A..50L} have presented a method to search for wobbles compatible with the presence of asteroid companions, leading to a set of 357 targets. The statistical selection of targets has been improved by \cite{2026arXiv260522702L}. However, the most promising candidates comprise those in common between the two analyses, and some of them are highlighted in the following section~\ref{sec:results}. For this study, we considered each binary candidate as a system rather than as an isolated body in order to properly distinguish and characterise the individual components and their mutual interactions. 
\\

To be able to assess the nature of these binary candidates, we propose using the stellar occultation technique. A combination of Gaia data to infer their binary nature and targeted follow-up stellar occultation observations was successfully applied to (4337) Arecibo. The companion was discovered through stellar occultation. \cite{2022MPBu...49....3G} and \cite{2023A&A...674A..12T} have shown that the astrometric wobble of a binary system with this configuration could be detected by Gaia thanks to the precision of the instruments. Subsequent stellar occultation observations confirmed the characteristics of the mutual orbit derived this way. This key method constitutes a proof of concept for our approach. Stellar occultations occur when three bodies align: an observer, an occulting Solar System object, and a background star. The occulting object passes in front of the star, creating a shadow detectable by the observer. As the object moves, the shadow also moves, creating an occultation path on the surface of the Earth. Analysis of the light curve emitted by the star using aperture photometry allows for the determination of key physical parameters of the object, such as the size, shape, orientation, and relative component geometry, with sub-kilometric accuracy. When recorded from multiple locations, each observation provides a chord across the object, which is a linear measurement of the object's limb. Stellar occultations also yield highly precise astrometric positions at the epoch of the event \citep{2019A&A...625A..43D,2020A&A...644A..40R} and offer a unique capability to probe the immediate environment of the object, enabling the detection and characterisation of satellites that would otherwise remain inaccessible. 

The efficiency and reliability of stellar occultation observations have been significantly enhanced by the involvement of coordinated amateur astronomer networks, which have demonstrated an exceptional ability to respond rapidly to observation campaigns and to provide dense chord coverage. This contribution has led to major results in recent years, including the precise characterisation of complex systems, such as ring systems for (10199) Chariklo, (50000) Quaoar, and (136108) Haumea \citep{2014Natur.508...72B,2023Natur.614..239M,2017Natur.550..219O}, and the discovery and follow-up observations of binary asteroids such as (90) Antiope \citep{2012LPICo1667.6427C}, (87) Sylvia \citep{2014Icar..239..118B}, and (216) Kleopatra, for its shape and multiple satellite system \citep{2011Icar..211.1022D,2022A&A...657A..76B,2021A&A...653A..57M}. Building on these experiences, our aim with this work is to conduct an observational study for each target identified by \cite{2024A&A...688A..50L} in order to confirm or refute their binary nature.
\\

This paper is organised as follows. In Sect.~\ref{sec:method} we describe the method used to predict, observe, and analyse the various targets through stellar occultations. Section~\ref{sec:results} presents five stellar occultation campaigns for four distinct MBAs. The various events presented highlight relevant challenges concerning the detection of satellites around intermediate-size (diameters <100 km) asteroid systems in the Solar System. The observation of (5044) Shestaka serves as an example of a large-scale campaign designed to probe its near environment. (35420) 1998 AG$_{6}$ and (206) Hersilia appear to be contact binary systems, or they exhibit close configurations that cannot be resolved without additional occultation observations. (1127) Mimi presents ambiguous results that can be enlightened by past stellar occultation data, and the (36882) 2000 SW$_{155}$ occultation event highlights the identification of a component that could be interpreted as a companion. A more global overview of the observed targets, together with associated descriptions of the most significant campaigns, is given in Section~\ref{sec:summary}. A discussion on the physical implications of the observations and comments on the current status of the campaigns are presented in Section~\ref{sec:improve}, while Section~\ref{sec:ccl} contains a summary of the work carried out and outlines future opportunities for the observation of binary asteroids.

\section{Occultation campaigns}
\label{sec:method}
\subsection{Predictions of stellar occultation events}

The success of a stellar occultation observation strongly depends on the timing accuracy and synchronisation of all observer as well as the astrometric accuracy of the occulted star and the occulting object. We use of the third release of the Gaia DR3 catalogue that provides accuracy up to one hundredth of a milliarcsecond (mas) for stars \citep{2023A&A...674A...1G}. Thus, most of the prediction uncertainty stems from the ephemeris of the occulting object, especially large ones. For each target in the GaiaMoons list, predictions and diffusion are made using the same method as the Lucky Star\footnote{\url{https://lesia.obspm.fr/lucky-star/index.php}} project (European Research Council funded) except that we use the most recent JPL ephemeris for asteroids orbit predictions : We perform a systematic search across the entire Earth for occultation candidate stars brighter than magnitude 14. A script  propagates the ephemeris of the mean system across the sky and uses the Gaia DR3 catalogue to predict upcoming occultations. We retrieve key information about the occulted stars and the occulting target. These predictions are then published on a dedicated website\footnote{\url{https://gaiamoons.imcce.fr/}} and shared with communities of both amateur and professional astronomers. 
\\

Each set of consecutive Gaia observations of the same object in a short time span (referred to as a “window”) provides a set of parameters associated with a potential companion, see details in \cite{2024A&A...688A..50L} and \cite{2023A&A...674A..12T}. Some objects are observed multiple times, producing several observation windows. When these different windows yield consistent sets of parameters — for instance, in terms of orbital period — this increases confidence in classifying the object as a binary system. These data are therefore more reliable and precise, making it possible to further constrain the orbit of the potential companion around the primary and to provide a prediction of the companion’s position relative to the primary during stellar occultation. This information is communicated directly to observers on a case-by-case basis as it is highly dependent on Gaia data for a given system. 

For each window we measure a period and an amplitude of the wobble, as projected on the sky, in the direction of highest accuracy of Gaia (the scan direction). Starting from these two components, and assuming a range of possible bulk densities (from 0.3 to 7 $g.cm^{-3}$), a simple model allows intervals of possible mass ratio (q) and and separation (S) of the components to be derived. For each system, three situations are possible: (1) no physically meaningful interval is found and the system is discarded; (2) one interval of possible q and S; (3) two intervals of possible q and S (see details in \cite{2024A&A...688A..50L}, 2026). In cases (2) and (3), diameters available in literature can be exploited to derive interval of possible absolute sizes for the two components.
For simplicity, in the following (including the related plots) we represent the allowed intervals by a mean value and a range (which should not be interpreted as an error bar). 

\subsection{Coordinated observations}

Among all predicted stellar occultation events, we marked some events as favourable according to different criteria. To select these events, we applied a hand-filtering process by focusing on areas with active observers and by taking into account the brightness of the star and the conditions of observation (elevation of the star, expected occultation duration). We rely on known international or local stellar occultation observer networks such as the International Occultation Timing Association (IOTA) with its different section in Europe (IOTA/ES) and East-Asia (IOTA/EA), the Trans-Tasman Occultation Alliance (TTOA) in Australia, and the Réseau d’Observateurs d’Astéroïdes DIspatchES (ROADIES) \citep{2022sf2a.conf..141D} in France. We then published These favourable events on the GaiaMoons website in a dedicated section. We regurarly communicate these events to the relevant communities in order to circulate calls for observations.
\\

When conditions are particularly favourable (expected good weather, bright star, observer availability), coordinated campaigns are organised. Observers equipped with mobile and fixed stations are distributed uniformly along the centrality (the expected central position of the shadow). This allows the uncertainty zone of the satellite to be systematically covered, thereby maximising the probability of detecting the satellite. In addition, we apply the method developed by \cite{2024A&A...688L..23L} to estimate the probable position - if it exists - of the satellite at the time of occultation using Gaia photometric data of the system (see detailed example on Section~\ref{results:shestaka}). This method allowed us to target a specific area especially when the number of observer is not sufficient to cover all the path uncertainties with a dense mesh. 
After the observations, raw data (and reduction performed by observers, when available) are collected via the OccultPortal platform\footnote{\url{https://occultationportal.org}} \citep{2022MNRAS.515.1346K}. Given the large number of events occurring daily and the diversity of the communities involved, there is a significant variability in the types of data received (video formats, FITS images, custom formats) and on the range of telescopes, from small portable ones to large fixed facilities.

\subsection{Data reduction and analysis}

When possible, we converted raw video files into FITS format using Siril \citep{2024JOSS....9.7242R} or a proprietary script based on Astropy \citep{2013A&A...558A..33A}, particularly for .avi, .adv and .ser files. We analysed these raw files using dynamic aperture photometry on the target star using - when possible - one or more nearby stars as calibrators to minimise noise and the impact of potential hazardous events such as clouds in the field of view. To do so we used the Package for the Reduction of Astronomical Images Automatically (PRAIA, \citealt{2011gfun.conf...85A,2023P&SS..23905816A}) and PyMovie \citep{2019JOA.....9d...9A} especially for other formats. As comprehensive tools, they allow dynamic tracking of the aperture size to maximise signal-to-noise ratio (S/N), among others. When possible, we used both tools to analyse the video to consolidate and validate the reliability of the reduction. No significant inconsistencies were found.
Similarly, for videos or fits captures, timing measurements are made through various means such as the Global Positioning System (GPS) or synchronisation via the user’s PC using the Network Time Protocol (NTP;  \citealt{1991ITCom..39.1482M}), which is employed by the majority of observers. For most systems, temporal delays of cameras and time recording systems were well characterised and were taken into account in the analysis.
\\

To organise our occultation analysis workflow we used the Python library Stellar Occultation Reduction and Analysis (SORA, \citealt{2022MNRAS.511.1167G}). Reduced light curves are then normalised around the mean value to be scaled to unity, and the immersion and emersion times are calculated using the classical reduced $\chi^2$ minimisation method (see details on \citealt{2022MNRAS.511.1167G}) between a synthetic light curve and the observed light curve. This method takes into account Fresnel diffraction effects, the sensor bandwidth (often visible optical CCD $\lambda = 550$~nm), the stellar diameter, and the image integration time \citep{2017Natur.550..219O,2011Natur.478..493S}. For this type of observation, one may aim to detect satellites of very small size, around one or two kilometres or grazing observations. Given an average semi-major axis of 2.8 AU (provided by the set of GaiaMoons targets) and a typical observation wavelength of $\lambda = 550$~nm, the typical Fresnel scale ($\lambda_{\rm F} = \sqrt{\lambda \Delta/2}$, where $\Delta$ is the geocentric distance) is 340~m. For objects with sizes larger than 2~km, the finite angular size of the star is generally negligible \citep{2024A&ARv..32....6S}. SORA models diffraction fringes of occulting objects as rectangles with infinite length \citep{2022MNRAS.511.1167G}. These simple models, 
originally designed for stellar occultations by large objects (TNOs, Trojans, natural satellites) are not adapted to grazing geometries, especially when diffraction effects are not negligible as seen in sections~\ref{results:AG6} (1998 AG$_6$) and \ref{results:SW155} (2000 SW$_{155}$). Diffraction effects depend on the star apparent velocity relative to the local limb of the object \citep{1987AJ.....93.1549R}. To assess this impact, we estimated the attack 
angle\footnote{A value of 90$^\circ$ indicates a star apparent velocity normal to the limb, a value of 0$^\circ$ indicates an apparent velocity parallel to the limb.} of each chord on the object and applied a projection on the shadow's normal velocity accordingly. This approach provides a more realistic modelling of the light curve and allows the corresponding drops to be confirmed within the error bars on Stations 1 and 3 for 1998 AG$_6$ and Station 1 for 2000 SW$_{155}$ (see Fig.~\ref{figA4}, (h) Oty\'n, (j) MSO, (s) Legionowo). 
\\

For each station, the goal is therefore to find the best compromise between a short exposure time that reduces the timing error on the event, and the S/N that is limited by the observing conditions, detectors and telescopes. After reducing the light curves, we model the observed limb as an ellipse using five parameters: the centres offsets ($f,g$) with respect to the associated JPL prediction ephemeris, the apparent equatorial radius a, the apparent oblateness $\epsilon = (a - b)/a $, b being the minor axis of the ellipse, and the position angle of the minor axis denoted by $PA$ and defined as the angle measured eastward from celestial north. The statistical significance of these fits are done using the reduced $\chi^2$ method. Unless stated otherwise, all quoted uncertainties correspond to $3\sigma$ confidence intervals, derived from the marginal probability distributions of each fitted parameter (i.e. the $99.73\%$ confidence level for a Gaussian distribution), and are considered independently of the other adjusted parameters. All positions are computed after correction from light deflection caused by the Sun. In the following we refer to surface-equivalent and volume-equivalent radii of a two-body system, which respectively stand in for the radius of a sphere having the same surface area\footnote{$R_{\rm equiv}^2 = R_{prim} ^2 + R_{sat} ^2$} and volume.

\section{Results for detection of potential binary systems}
\label{sec:results}

\begin{center}
    \begin{table*}[ht!]
        \small
        \caption{Occulted star parameters for each detailed event.}
        \label{tableA1}
        \begin{tabular}{c c c c c c}
            \hline\hline
            Name & Date at closest approach UTC & Gaia DR3 identifier & Right Ascension$^{(ii)}$ & Declination$^{(i)}$ & G mag$^{(ii)}$\\
                  & yyyy-mm-dd hh:mm:ss & source identifier & hh mm ss.sss (mas) & dd mm ss.sss (mas) & \\
            \hline
            Shestaka & 2024-10-23 18:55:15.0 & 6842345973017889152 & $+21^\mathrm{h}\,30^\mathrm{m}\,27.31532^\mathrm{s}\,(0.3)$ & $-14$° $18'$ $30.40563"$ (0.1) & 10.5\\
            AG6 & 2024-07-17 01:43:32.0 & 4204153068710661760 & $+19^\mathrm{h}\,09^\mathrm{m}\,02.16451^\mathrm{s}\,(0.2)$ & $-09^\circ\,02'\,51.93763"\,(0.2)$ & 11.7\\
            Hersilia & 2026-01-12 20:32:14 & 3365619562371972608 & $+06^\mathrm{h}\,46^\mathrm{m}\,42.20411^\mathrm{s}\,(0.2)$ & $+19^\circ\,09'\,39.0438"\,(0.1)$ & 13.1\\
            Mimi & 2025-02-26 19:29:51.0 & 3366345721081820288 & $+06^\mathrm{h}\,58^\mathrm{m}\,20.43935^\mathrm{s}\,(0.2)$ & $+21^\circ\,20'\,14.8745"\,(0.2) $ & 12.9\\
            Mimi & 2025-02-16 01:28:56.0 & 3365895054454652544 & $+06^\mathrm{h}\,57^\mathrm{m}\,02.18043^\mathrm{s}\,(0.3)$ & $+19^\circ\,54'\,46.0955"\,(0.2)$ & 10.4\\
            Mimi$^{(*)}$ & 2023-07-21 14:00:07.4 & 4089152501854608256 & $+19^\mathrm{h}\,07^\mathrm{m}\,01.00027^\mathrm{s}\,(0.2)$ & $-15^\circ\,21'\,17.4734"\,(0.2)$ & 14.6\\
            Mimi$^{(*)}$ & 2019-11-19 10:12:22.6 & 6805959353684555520 & $+20^\mathrm{h}\,54^\mathrm{m}\,23.53859^\mathrm{s}\,(0.1)$ & $-23^\circ\,58'\,53.6112"\,(0.1)$ & 11.3\\
            SW155 & 2025-08-29 22:57:31.0 & 4227288133359092224 & $+20^\mathrm{h}\,53^\mathrm{m}\,05.9784^\mathrm{s}\,(0.3)$ & $-00^\circ\,33'\,59.0585"\,(0.2)$ & 11.2\\
            
            \hline
            \hline
        \end{tabular}
        \tablefoot{
        \tablefoottext{i}{Star astrometric position include proper motion and parallax and are given in ICRF (J2000).}
        \tablefoottext{ii}{Visual magnitude retrieved from Gaia DR3 catalogue release \citep{2023A&A...674A...1G}}
        \tablefoottext{*}{Past occultation where timing were retrieved from the Planetary Data Center database \citep{2020MNRAS.499.4570H}.}
        }
    \end{table*}
\end{center}

\subsection{(5044) Shestaka on October 23, 2024}
\label{results:shestaka}

\begin{figure*}[ht!]
\centering
\includegraphics[width=\hsize]{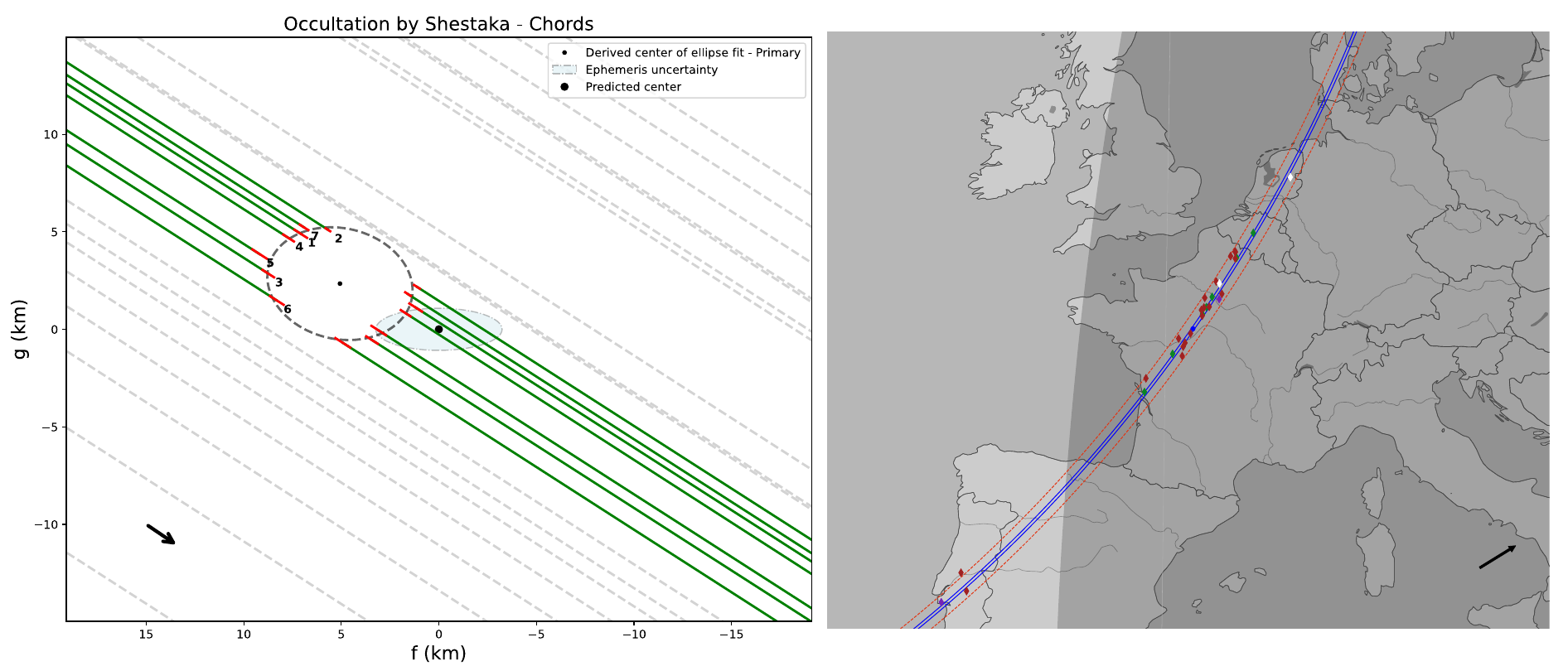}
  \caption{
  Left: Results of the stellar occultations by (5044) Shestaka on 2024 October 23. The positive chords are displayed in green, where the interruptions correspond to the star disappearances. Ingress and egress uncertainties are displayed in red segments with $1\sigma$ uncertainty. Displayed numbers in black correspond to stations (see Table~\ref{tab:5044}). The dashed lines in light grey are the negative chords. The large black dot at  the origin represents Shestaka's centre position, with its $3\sigma$ uncertainties domain displayed in light blue area (see Table~\ref{table1} for fit details). The elliptical fit solution is displayed with a  dashed black line, where the smaller black dot represents the derived centre of the primary. The fit shows a large shift between the main object and the prediction. North is up, east is left, and the direction of the object in the sky plane is given by the black arrow.
  Right: Reconstructed Shestaka's shadow track over Western Europe. Blue lines represent the limit of the primary shadow path, and the dotted red lines represent the area of uncertainty for the  position of the satellite. The observing stations are displayed with the following colour-code: red - negative, green - positive, white - overcast, purple - technical issues. The black arrow shows the direction of motion of the shadow. The dark grey region corresponds to astronomical night, while the light grey region corresponds to astronomical twilight.}
     \label{fig2}
\end{figure*}

The 2024 October 23 stellar occultation by (5044) Shestaka system\footnote{\url{https://gaiamoons.imcce.fr/occ.php?p=28421}} was identified as a favourable one among the events predicted for 2024.  It was easily observable from Western Europe, particularly in France and Portugal (see Fig~\ref{fig2}) and involved a bright (G$_{\rm mag}$=10.5) star, see details in Table~\ref{tableA1}. Shestaka's shadow swept the Earth's surface from 18:47:45 to 19:02:42 UT, with a geocentric closest approach at 18:55:15 UT. A total of 30 stations were involved in the campaign, and seven of them detected a positive chord. The circumstances of observations are given in Table~\ref{tab:5044} and the occultation light curves are displayed in Figure~\ref{figA4}~(a-g). We note that these detections were obtained using small telescopes with diameter ranging from 15~cm to 35~cm. A total of 19 negative observations were recorded as well, which allowed the size and position of the asteroid to be efficiently constrained and its direct surrounding environment to be probed with great accuracy (see Fig~\ref{fig2}).

\begin{table}[ht!]
    
    \caption{Ellipse fit details for 5044 Shestaka.}
    \renewcommand{\arraystretch}{1.2}
    \begin{tabular}{l r}
        \hline\hline
        Apparent semi-major axis & $a$ = ($ 3.6 \pm 0.2 $) km\\
        Apparent oblateness & $\epsilon$ = $ 0.20 \pm 0.03 $\\
        Position angle & $PA$ = ($ 162.0 \pm 5.0 $) deg\\
        Equivalent radius\textsuperscript{(a)} & $R_{\rm equiv}$ = ($ 3.2 \pm 0.2 $) km\\
        Date (UT) &  2024-10-23 18:55:14.32 \\
        \hline
        Right Ascension\textsuperscript{(b)} & $+21^\mathrm{h}$ $30^\mathrm{m}$ $27.17209^\mathrm{s}$~(0.4) mas \\
        Declination & $-14$° $18'$ $26.7180"$~(0.4)  mas \\
        \hline\hline
    \end{tabular}
    
    \label{table1}
    \tablefoot{
    Astrometric positions are given in geocentric International Celestial Reference Frame (ICRF; J2000) at the given date referring to the time of closest approach (TCA) which is the moment when the apparent distance between the asteroid and the star, as seen from the geocentre, is at its minimum. Reference prediction is given by \#JPL67 ephemeris. The scale in the asteroid's orbit is 1~mas = 1.08~km.
        \tablefoottext{a}{The surface-equivalent radius is given by $R_{\rm{equiv}} = a\sqrt{1-\epsilon}$.}
        \tablefoottext{b}{The astrometric position is computed using the associated centre of the ellipse (see Fig~\ref{fig2}). Centre offsets are $(f,g) = (6.3, 2.7)$~km from \#JPL67 ephemeris prediction.}}
\end{table}

According to \cite{2024A&A...688A..50L}, the two predicted solutions for separation were $17.7 \pm 4.4$~km and $16.6\pm 3.8$~km in the celestial plane, we took the upper limit of 22.2~km as maximum uncertainty region of the presence of the satellite. Additionally, we applied the method developed by \cite{2024A&A...688L..23L} using previous photometric data of this system to estimate the probability of presence of the potential satellite at the time of closest approach (TCA). This method did not lower the initial estimated separation between the satellite and the primary to constrain of search area for observer. The volume-equivalent diameter available in the literature for the primary was $6.4 \pm 0.2$~km as reported from the Wide-field Infrared Survey Explorer (WISE, \citealt{2022PSJ.....3...30M}). The left panel of Figure~\ref{fig2} displays the positive and negative chords projected onto the celestial plane at TCA and the elliptical limb fit obtained from the $\chi^2$ minimisation fit method. The parameters of the fit are given in Table~\ref{table1}. The fit reveals a fairly regular and elongated shape. We estimate the surface-equivalent diameter of $6.4 \pm 0.4$~km which is consistent with the WISE estimation within the uncertainties. 
The elliptical limb model provides an on-sky offset of 5.2~mas (5.6~km in the asteroid's orbit) for Shestaka with respect to the prediction \#JPL67. This is larger than estimated uncertainties of the prediction  3.2~mas (3.4~km, $3\sigma$). This relatively large offset may be interpreted as caused by a satellite even should be regarded with caution as the position is given with respect to a single prediction ephemeris.
\\

Different stations scanning the area around the object used average exposure times of 50~ms (Table~\ref{tab:5044}), it should have revealed a drop caused by an object with a diameter of at least 450~m, considering the shadow velocity of 9~km.s$^{-1}$. The predicted satellite diameter was estimated between 2~km and 5~km \citep{2024A&A...688A..50L} and should have been observable within these positions. Therefore, the negative (and positives outside of the primary occultation region) chords not only constrain the size and position of the primary, but also can exclude regions where the putative satellite could be. This critical probing will help future observations for this specific system. We note that some areas are not covered by our observations, due to technical issues or weather conditions, see the region not covered in the northeast corner of Figure~\ref{fig2}.

\subsection{(35420) 1998 AG$_6$ on July 17, 2024}
\label{results:AG6}

The asteroid (35420) 1998 AG$_6$ (hereafter denoted AG6 for brievity) is also an outer main belt object with a volume-equivalent diameter of $6.6 \pm 0.1$ km according to \cite{2012ApJ...759L...8M} using Near-Earth Object Wide-field Infrared Survey Explorer (NEOWISE) thermal data. It was observed for the first time by stellar occultation on July 17, 2024\footnote{\url{https://gaiamoons.imcce.fr/occ.php?p=28338}}, by seven stations in Europe, including three positive chords in France and Poland (see Fig~\ref{figA6}), these chords are displayed in Figure~\ref{figA4}~(h-j). The occulted star with magnitude $G_{\rm mag} = 11.7$ (see Table~\ref{tableA1}) and the shadow swept the Earth's surface between 01:36:57 and 01:50:04 UT. Observing stations information are available in Tab~\ref{ tab:35420 }.

\begin{table}[ht!]
    \caption{Ellipse and shape model parameters for stellar occultations by AG6.}
    \renewcommand{\arraystretch}{1.2}
    \begin{tabular}{l r}
        \hline\hline
        \multicolumn{2}{c}{(a) Ellipse fit}\\
        \hline\hline
        \multicolumn{2}{c}{AG6.1 - Primary} \\
        \hline
        Equivalent radius & $R_{\rm equiv} = ( 2.4 \pm 1.9 )$~km\\
        Right Ascension$^{(ii)}$ & $+19^\mathrm{h} 09^\mathrm{m} 2.09444^\mathrm{s}$ (0.75)~mas \\
        Declination & $-09^\circ\,02'\,49.1679"$ (1.47)~mas \\
        \hline
        \multicolumn{2}{c}{AG6.2 - Satellite} \\
        \hline
        Equivalent radius & $R_{\rm equiv} = ( 2.0 \pm 1.5 )$~km\\
        Right Ascension$^{(i)}$ & $+19^\mathrm{h} 09^\mathrm{m} 2.09463^\mathrm{s}$ (1.47)~mas\\
        Declination & $-09^\circ\,02'\,49.1679"$ (0.76)~mas \\
        \hline\hline
        \multicolumn{2}{c}{(b) Shape model} \\
        \hline\hline
        Sub-observer coordinates  & $(\lambda_{SEP},\,\beta_{SEP}) = ( 200.0 ,\, 23.3)\,^\circ$\\
        Pole position angle & $ PA= 173.1 \,^\circ$\\
        Pole aperture angle & $AA = 23.4 \,^\circ$\\
        Equivalent radius & $R_{\rm equiv} = ( 3.3 \pm 0.3 )$~km\\
        Right Ascension$^{(iii)}$ &  $+19^\mathrm{h} 09^\mathrm{m} 2.09450^\mathrm{s}$ (1.47)~mas \\
        Declination & $-09^\circ\,02'\,49.1683"$ (1.43)~mas \\
        \hline
    \end{tabular}
    
    \label{tableb1}
    \tablefoot{Astrometric positions are given in geocentric ICRF at TCA 2024-07-17 01:43:32.06. Results are given with respect to the prediction JPL\#59. The scale in the asteroid's orbit is 1~mas = 1.38~km. $AA$  is  the  angle  between  the  rotation  pole  of the asteroid and the line of sight from the geocentre.
    \tablefoottext{i}{centre of ellipse in the projected sky plane $(f,g) = (0.1, -2.1)$~km.}
    \tablefoottext{ii}{$(f,g) = (-3.8,-2.2)$~km.}
    \tablefoottext{iii}{$(f,g) = (-2.5,-2.6)$~km (see Fig~\ref{fig4}).}
    }
\end{table}

The distribution of the chords (with Stations 1 and 3 located near the external part of the limb in particular) combined with the total number of observers did not allow the shape of the object to be tightly constrained. Yet, the negative chords recorded by Stations~4 and 5 help constrain the position of the body on the northern side. Several interesting features can be discussed; Station~1, in particular, was able to measure the event with a very short exposure time of 40~ms (see Table~\ref{ tab:35420 }) and a negligible dead time and therefore recorded two successive drops visually confirmed, lasting respectively 270~ms and 130~ms and separated by 160~ms (Fig~\ref{figA4}~(h) Oty\'n and (j) MSO). The case of a double star is excluded as the depth of the first drop is $\sim 99\%$, consistent with the predicted magnitude drop of 6.1 (99.6\%).

The second drop observed at Station~1 has a depth of around 60\% and stands a grazing observation, which can be explained by the predominance of diffraction effects and the fact that the apparent size of the star at the object's distance (400~m at 1.89~AU) is no longer negligible for the drop depth. For this event, the Fresnel scale is $\lambda_{\rm F} = 279$~m, the angular diameter of the star projected at the body is 400~m. Station~1 passed near the edge of the object. In this geometry, diffraction effects cause a partial drop of the stellar flux. At the same time, diffraction effects widen the dip in the light curve, leading to an overestimation of the occulter size. These effects, convolved with the exposure time, reduce further the stellar drop (see \cite{1987AJ.....93.1549R}). 

With the same reasoning sequence, Station 3 gives a similar result at a corresponding time, therefore reinforcing the robustness of our approach. This way, two drops with durations of 200~ms and 50~ms are identified, which is consistent with the observation at Station 1, the two stations being close once projected onto the celestial plane. Finally, Station 2 recorded a single drop of 0.6~s, which is more than twice the duration of the main drops recorded at Stations~1 and 3. The final shape of the object reconstructed in this way cannot be approximated by a simple ellipse. Two hypotheses can therefore explain the profile. Firstly, AG6 could be a binary asteroid whose two components were observed during an ongoing mutual event, preventing a firm conclusion on its nature. We note that Station~2 had a cycle time of 0.1~s, which could mask some details such as a small separation between the two bodies. The shape of the drop suggests that it is possible that the final drop is a blend of two individual drops. Secondly, AG6 could be an irregular contact binary.

\begin{figure*}[ht!]
\centering
\includegraphics[width=0.49\hsize]{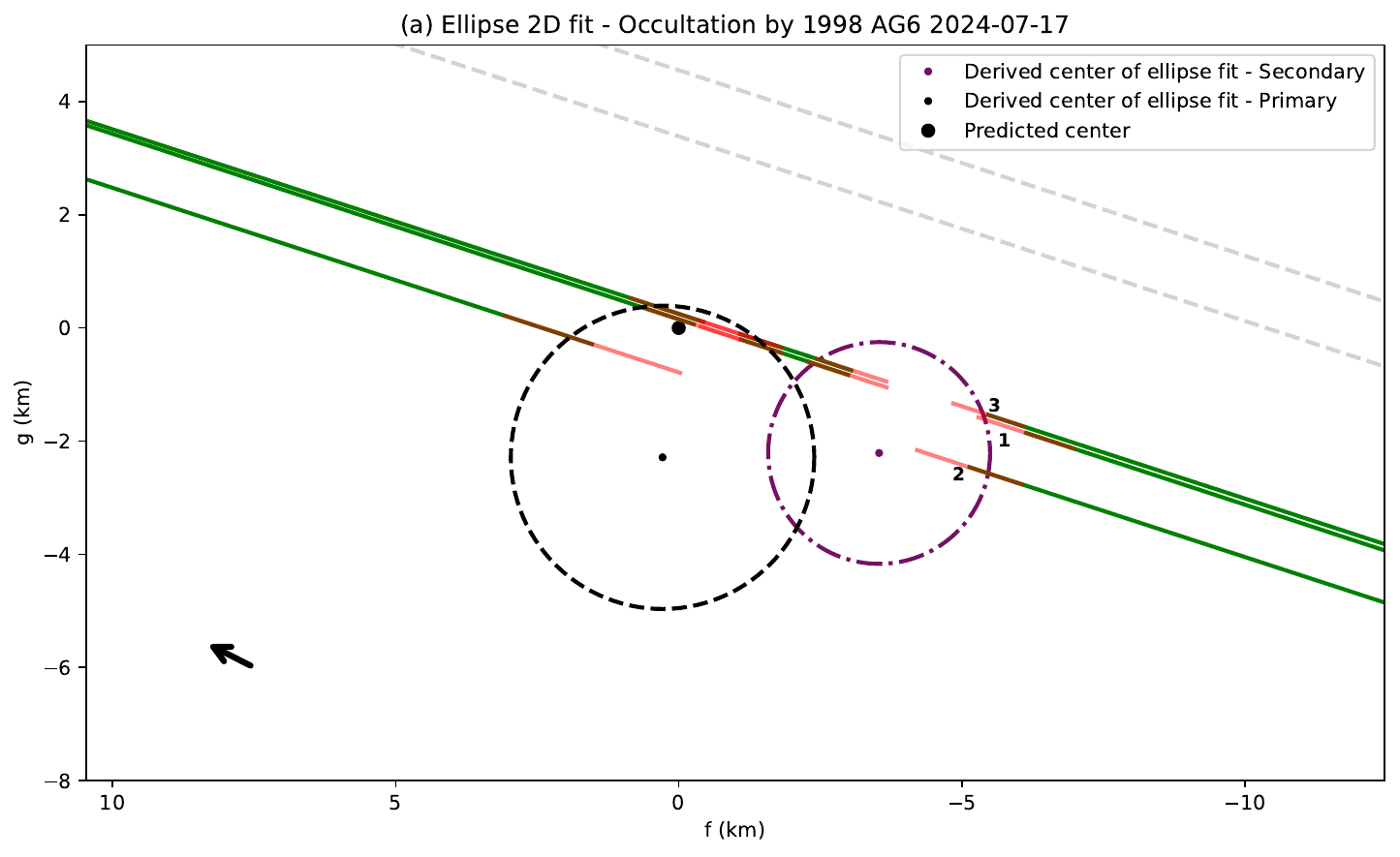}
\includegraphics[width=0.49\hsize]{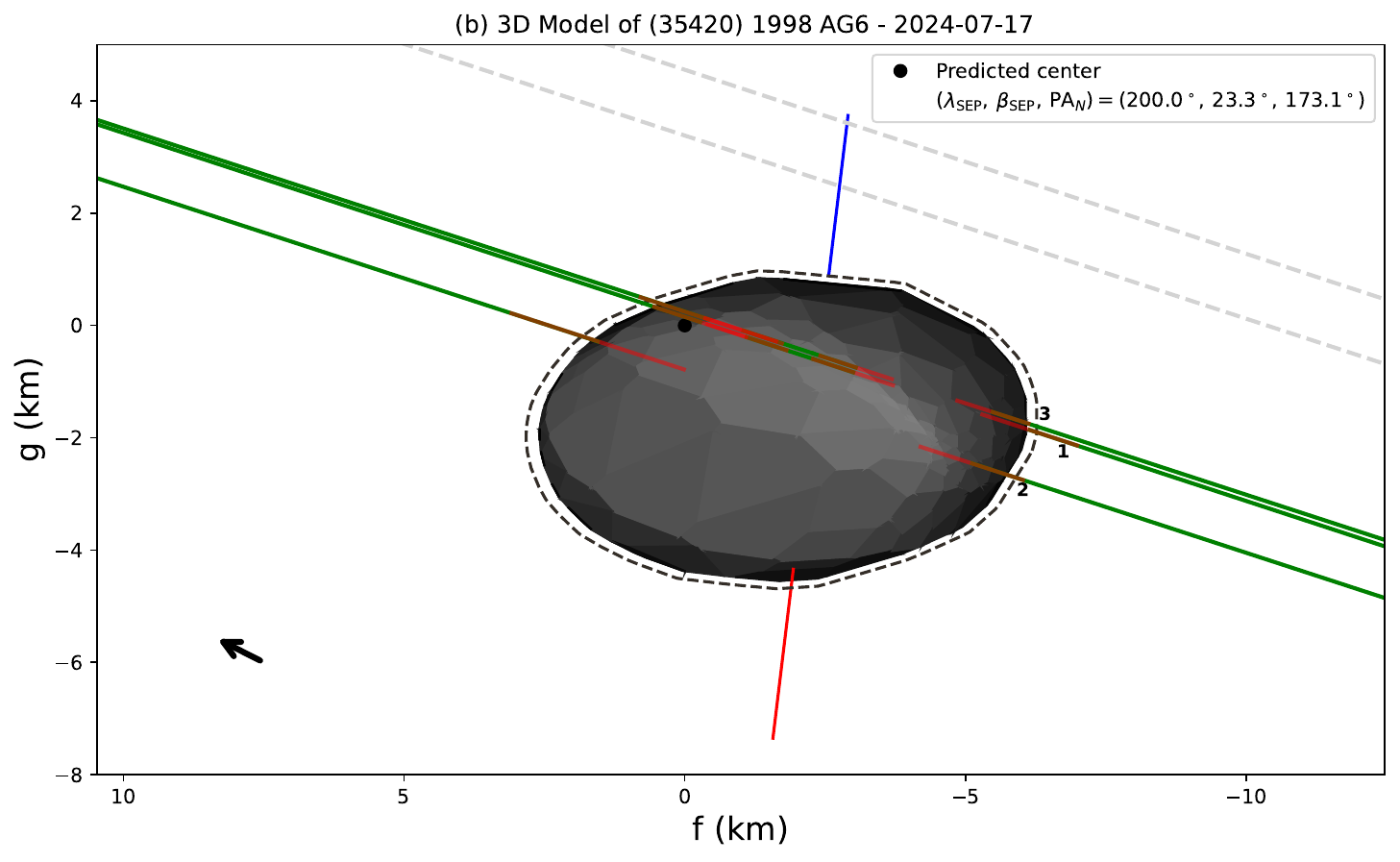}
  \caption{Best-fit sky plane projections of AG6 stellar occultation event of 2024 July 17. The chords are numbered and refer to the data in Table~\ref{ tab:35420 }. Uncertainties are displayed in semi-transparent red. Legends are the same as in Figure~\ref{fig2}. Fit details are given in Table~\ref{tableb1} Left: Result of the two-body fit. The satellite is shown in purple, and the primary is in black. The centre of the prediction is indicated by a thick black dot. Right: Result of the fit using the shape model 11119 from DAMIT. The pole is represented by the red (north pole) and blue (south pole) axes. The green area drawn by chords 1 and 3 are the separations between the drops; this space stands for a high constraint in the shape of the object.}
     \label{fig4}
\end{figure*}

The asteroid (35420) 1998 AG$_6$ is part of the DAMIT database \citep{2010A&A...513A..46D} and therefore has a shape model derived from light-curve inversion with Gaia photometric observations (model number 11119, \citealt{2023A&A...675A..24D}). The reliability of the model should be treated with caution, as the light curve inversion technique produces convex models that cannot faithfully reproduce the small-scale irregularities of asteroid shapes. In addition, the shape model is assigned a quality flag of 1 in the DAMIT database, indicating that the solution is based primarily on sparse data. Such models are expected to be coarse, with potential significant uncertainties in both the derived shape and the pole orientation. We use the asteroid shape model to perform a least-squares fit using SORA in order to constrain the rotation of the body. For the fit, we take into account the immersion times of the first drop and the emersion times for Stations 1 and 3 referred as outer times in the following. As the pole solutions are poorly constrained, we restricted the fit to the rotation of the object since the observed chords constrain the shape model too weakly to allow for a reliable fit. As depicted in Figure~\ref{fig4}~(b) the shape model then reproduces satisfactorily the event described by the outer times, but the observations imply the existence of a significant depression at the surface of the body in order to explain the two drops.

\begin{figure}[ht!]
\centering
\includegraphics[width=0.8\hsize]{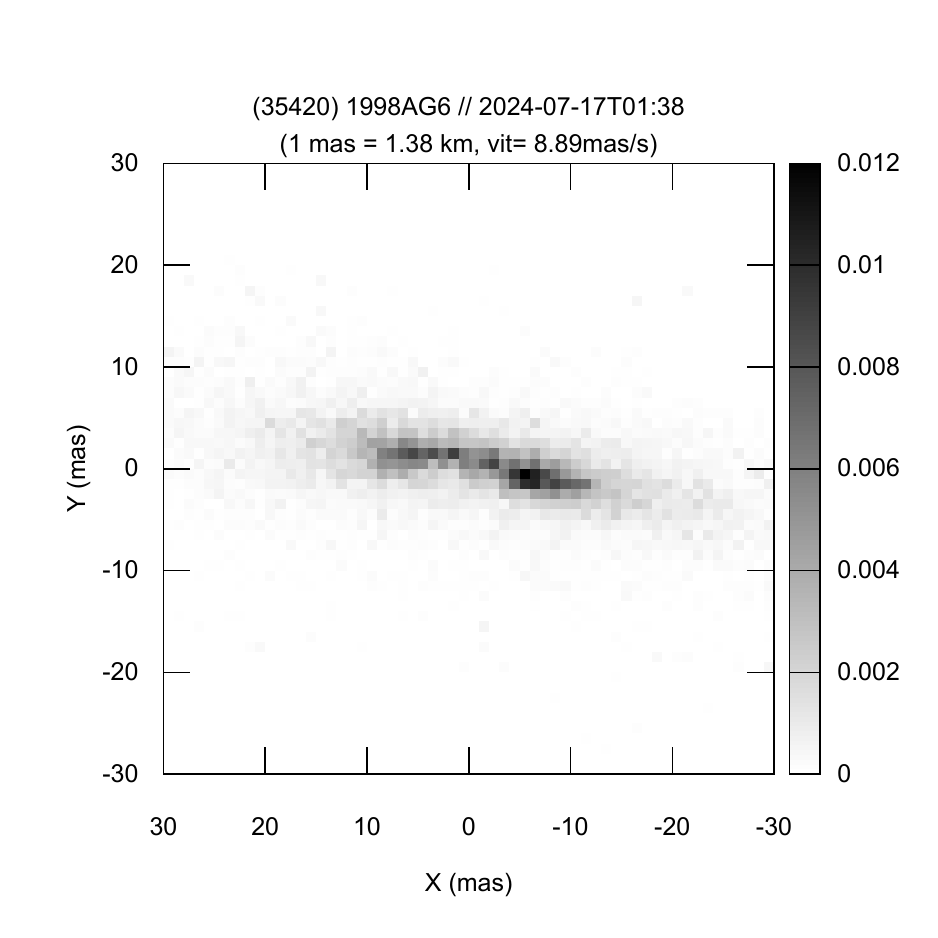}
  \caption{Probability estimation of the position of the satellite of AG6 based on Gaia photometric data using the method described by \cite{2024A&A...688L..23L}. The scale is the same as Figure~\ref{fig4}. The grid is composed of 2~km x 2~km squares with integrated probability of presence of the satellite. This distribution highlights two opposite areas around the centrality.}
     \label{fig:prediction}
\end{figure}

To model the observation more faithfully, we can instead perform a fit using two independent circles where the results are displayed in Figure~\ref{fig4}~(a). The characteristics and positions derived for the two hypotheses are provided in Table~\ref{tableb1}. The uncertainties associated with the drops imply important uncertainties in the size and positions of the two objects thus derived. This is reflected in the surface-equivalent radii of the two bodies $R_{\rm prim} = 2.7 \pm 1.9$~km and $R_{\rm sat} = 2.0 \pm 1.5 $~km. The body with a bigger mean diameter is arbitrarily considered as the primary and the smaller one as the satellite (see Fig~\ref{fig4}~a). 
\\ 

Considering the large uncertainties in the diameter, calculating the mass ratio of the system did not allow us to distinguish between the two solutions obtained by \cite{2024A&A...688A..50L}. With the present data, both solutions remain possible ($q_1 = 0.008 \pm 0.003$ and $q_2 = 0.672 \pm 0.056$). However, we note that the optimal solution taking mean values (see Fig~\ref{fig4}~a and Table~\ref{tableb1}) the mass ratio (0.578) is close to the second interval $q_2$. Additionally, the position of the two bodies is coherent with the estimation of the satellite position using available Gaia astrometry for this object \citep{2016A&A...595A...1G} (see Fig~\ref{fig:prediction}, \citealt{2024A&A...688L..23L}), as the two bodies were predicted to be aligned and within the central path. Additionally, the cumulative surface-equivalent diameter of the two ellipses, $D_{\rm equiv} = 6.25 \pm 3.5$~km. This value is consistent with WISE thermal data. In short, the current occultation data did not allow us to unambiguously determine the nature of AG6 though it permits to say that the DAMIT convex shape is not reproducing well the global shape. Both hypotheses of a binary system observed during a mutual event and an irregular contact binary remain plausible scenarios, which can only be discriminated through future stellar occultations and complementary photometric or high-time-resolution observations.

\subsection{(206) Hersilia on January 12, 2026}
\label{results:Hersilia}

(206) Hersilia is an MBA that has been selected as binary candidate after the new selection method presented in \cite{2026arXiv260522702L}. The object has been observed by occultation five times between December 2025 and February 2026. Two positive chords were recorded in Spain on December 19, 2025 (see more details on Section~\ref{sec:summary}), a single positive in Japan on January 1, 2026 and a double-positive chord has been measured on January 12, 2026. This event\footnote{\url{https://gaiamoons.imcce.fr/occ.php?p=90562}} is the main focus of this section, the shadow’s trail swept across the surface between 20:23:05 and 20:41:13 UT and the observation led to one positive (Fig~\ref{figA4}~(k)) and one negative observation in Spain. The general circumstances of the event are presented in Table~\ref{tableA1}. 

The occulted star magnitude was 13.1 (Table~\ref{tableA1}) and the apparent magnitude of the object was expected to be 12.5, leading to a magnitude drop of 0.5 (36.9\% of drop). The positive chord measured in station 1 (see Figure~\ref{figA4}~(k)) highlights two drops with equal depth around 27\%, the hypothesis of a double star with equal brightness (as indicated by the two drops) is excluded as the drops would be around 17\% in this case. The discrepancy between the expected and measured depth may come from different reasons: the apparent magnitude of the object may be poorly estimated, the observation is a graze as suggested by the position of the observer related to the prediction (see Figure~\ref{figA6}) or a combination of both. The apparent magnitude used does not take into account the shape model of the object \citep{2024A&A...687A..38C} and the available shape model comes from sparse data from Gaia and Lowell Obs sky survey 1998-2012 photometric measurement (quality flag 1, \citealt{2019A&A...631A...2D}). Moreover, the interpretation of this double drop is similar to AG6, a contact binary with two components or a satellite close to the primary body at occultation time. Additionally, the Fresnel scale of this event is 262 m, the velocity of the shadow was 11.14~km.s$^{-1}$ with 0.1s of total cycle time, giving a motion during each frame of 1114 metres per exposure. This would mean that the attack angle of the chords would need to be at least 75° normal to the limb to expect gradual drop or rise in the light curves. As the system is constrained by a single chord, only a few interpretations can be made. 
\\

\begin{figure}[ht!]
\centering
\includegraphics[width=0.9\hsize]{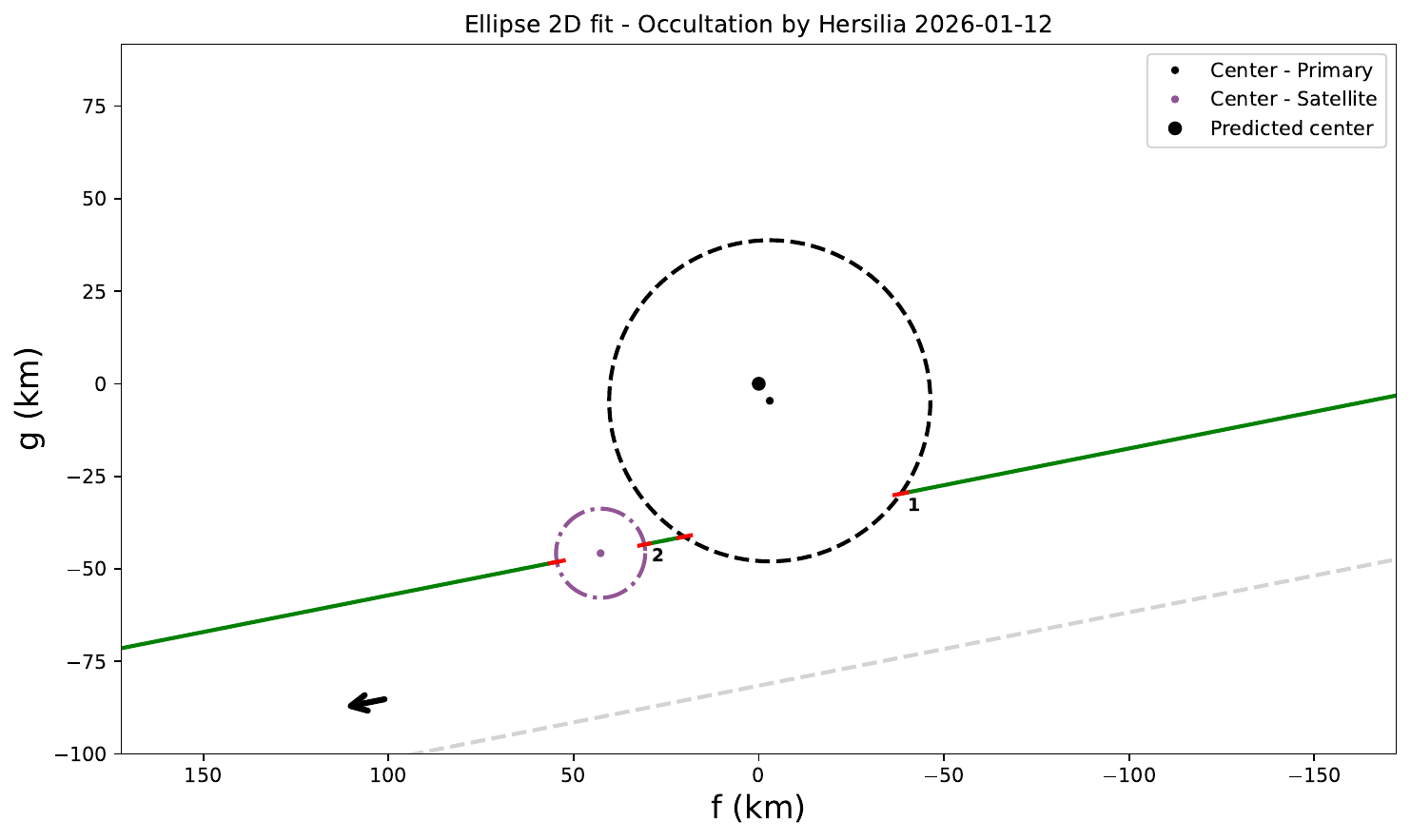}
\includegraphics[width=0.9\hsize]{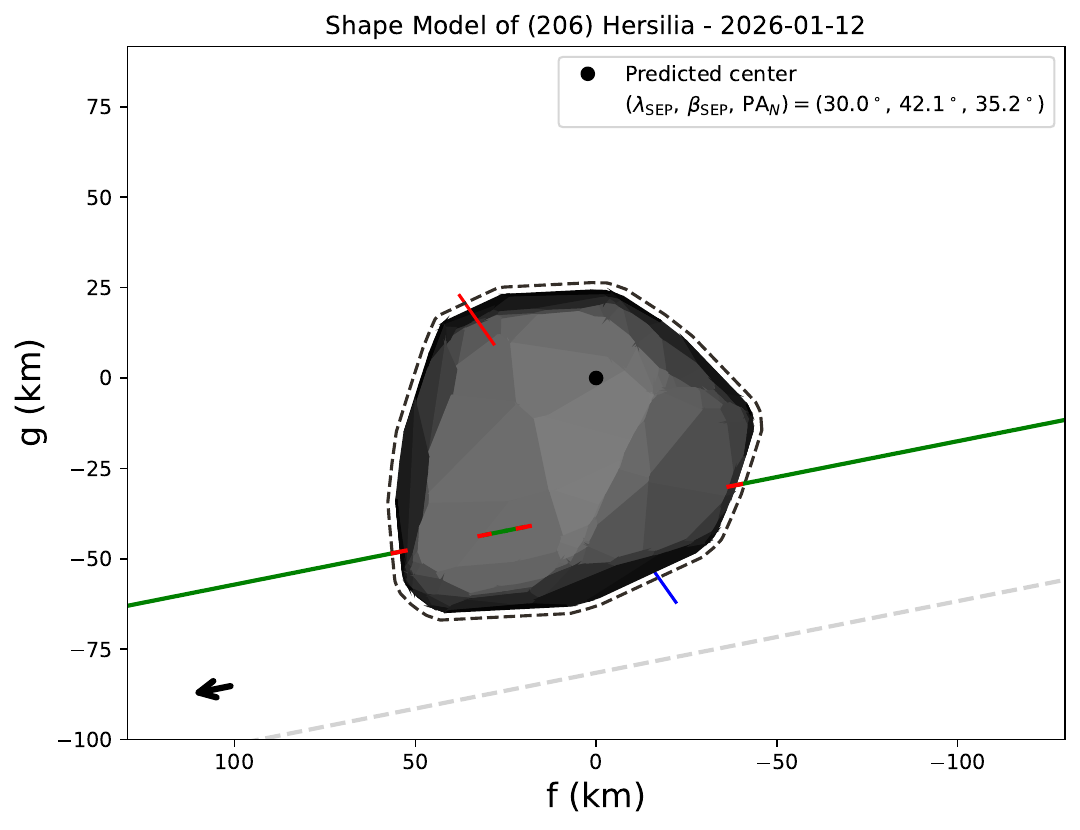}
  \caption{Best-fit sky plane projections of Hersilia stellar occultation event of January 1, 2026. Associated fit results are given in Table~\ref{tableHersilia}. The legend for the shape and ellipse fit are the same as in Figure~\ref{fig4}. Top: Result of the two-body fit. (Bottom) Result of the fit using the shape model 3217 from DAMIT. The green zone between the two drops highlight a concave feature of separation between the bodies.}
     \label{figHersilia}
\end{figure}

Considering the hypothesis that the object is a single contact binary object, we performed a limb fit (Tab~\ref{tableHersilia}) using the available shape model (n°3124) on DAMIT (see Figure~\ref{figHersilia}~down) with a size of 91.7 km \citep{2022PSJ.....3...56H}. The observation could invalidate the shape model and reveal local concavity at the edge. With the given orientation the observation could be a graze. For the binary system hypothesis, one can fit a circle at the middle of the second drop that would lead to a satellite with a minimum diameter of 17~km (Tab~\ref{tableHersilia}). The maximum size of the primary object is set to 96~km as the latest measured size with the upper uncertainties. These extreme values leads to a minimum mass-ratio 0.0064. As this value is considered as the bare minimum value for a binary hypothesis, we can discard the lower limit mass-ratio interval computed by \cite{2026arXiv260522702L} $q_1 = 0.0011 \pm 0.0001$. The size of the satellite is then constrained by the maximum size of the primary. 
The results of the observation from January 12 motivated observers to perform observations on January 17 and February 15, 2026 afterwards, without further information due to bad weather conditions. Other observations will be necessary to properly characterize the nature of this system and discriminate between a binary system or a contact binary.

\begin{table}[ht!]
    \caption{Ellipse and shape model parameters for stellar occultations by Hersilia.}
    \renewcommand{\arraystretch}{1.2}
    \begin{tabular}{l r}
        \hline\hline
        \multicolumn{2}{c}{(a) Ellipse fit}\\
        \hline\hline
        \multicolumn{2}{c}{Hersilia.1 - Primary} \\
        \hline
        Equivalent radius & $R_{\rm equiv} = 45.2^{+2.8}_{-10.9}$~km\\
        Right Ascension$^{(ii)}$ & $+06^\mathrm{h} 46^\mathrm{m} 42.22071^\mathrm{s}$ (6.25)~mas \\
        Declination & $+19^\circ\,09'\,40.4528"$ (8.28)~mas \\
        \hline
        \multicolumn{2}{c}{Hersilia.2 - Satellite} \\
        \hline
        Equivalent radius & $R_{\rm equiv} = 12.0^{+36.0}_{-3.5}$~km\\
        Right Ascension$^{(i)}$ & $+06^\mathrm{h} 46^\mathrm{m} 42.22361^\mathrm{s}$ (11.36)~mas \\
        Declination & $+19^\circ\,09'\,40.4280"$ (27.36)~mas\\
        \hline\hline
        \multicolumn{2}{c}{(b) Shape model} \\
        \hline\hline
        Sub-observer coordinates  & $(\lambda_{SEP},\,\beta_{SEP}) = ( 197.5 ,\, 23.3)\,^\circ$\\
        Pole position angle & $ PA= 173.12 \,^\circ$\\
        Pole aperture angle & $AA = 23.32 \,^\circ$\\
        Equivalent radius & $R_{\rm equiv} = ( 45.8 \pm 4.5 )$~km\\
        Right Ascension$^{(iii)}$ &  $+06^\mathrm{h} 46^\mathrm{m} 42.22126^\mathrm{s}$ (3.93)~mas \\
        Declination & $+19^\circ\,09'\,40.4392"$ (5.36)~mas \\
        \hline
    \end{tabular}
    \label{tableHersilia}
    \tablefoot{Astrometric positions are given in geocentric ICRF at TCA 2026-01-12 20:32:14.20. Results are given with respect to the prediction JPL\#134. The scale in the asteroid's orbit is 1~mas = 1.21~km. Mean diameter values are the same as the one projected in Figure~\ref{figHersilia}. 
    \tablefoottext{i}{centre of ellipse in the projected sky plane in km (see Fig~\ref{fig4}) $(f,g) = (-2.9, -1.9)$~km.}
    \tablefoottext{ii}{$(f,g) = (42.9,-45.7)$~km.}
    \tablefoottext{iii}{$(f,g) = (7.8,-19.5)$~km.}
    }
\end{table}

\subsection{(1127) Mimi on February 2025}
\label{results:mimi}

\begin{figure*}
\centering
\includegraphics[width=0.49\textwidth]{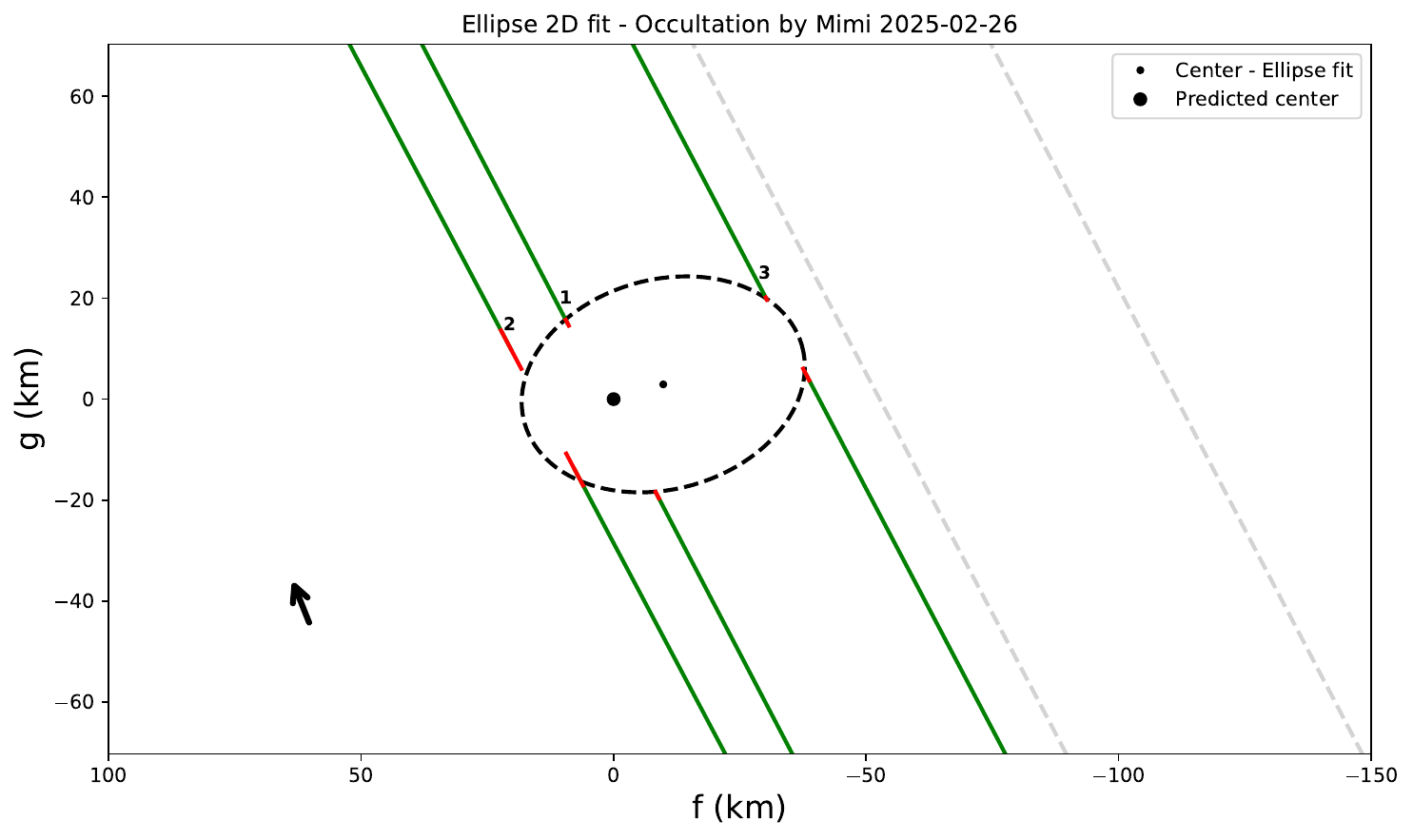}
\includegraphics[width=0.49\textwidth]{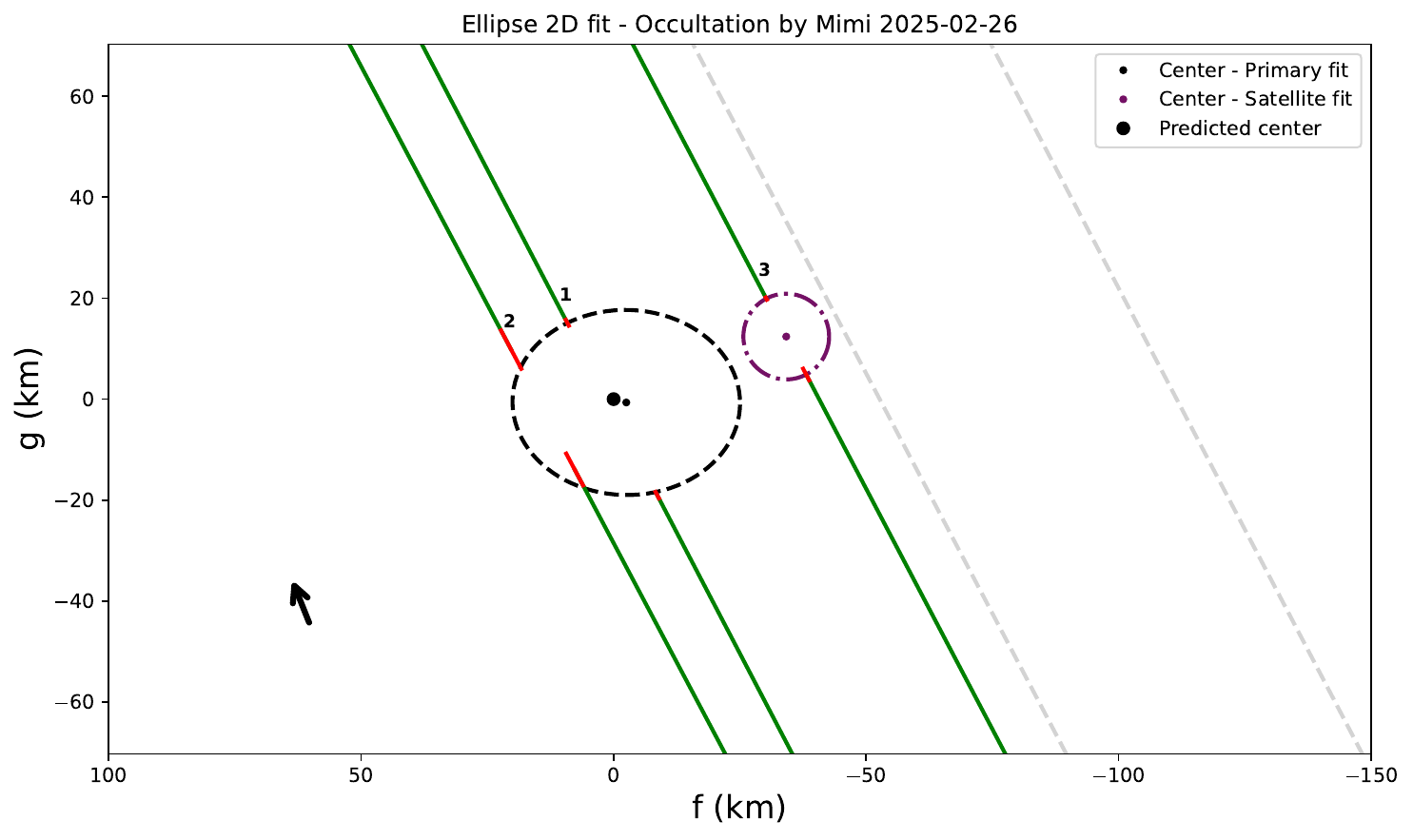}
  \caption{Left: Result of ellipse fit for the hypothesis of non-detection. Right: Result of ellipse fits for the main body and a satellite after stellar occultation event by (1127) Mimi on February 26, 2025. The primary body is represented by the dashed black ellipse. The satellite is in dash-dotted purple circle, and its diameter corresponds to the lower limit value $D_{\rm sat} = 17$~km. Details of the fit are given in Table~\ref{table2}. Legends are the same as in Figure~\ref{fig2}. Station details are available in Tab~\ref{tab:1127} (Appendix C) with corresponding numbers as displayed in the figure.}
     \label{fig3}
\end{figure*}

(1127) Mimi is an MBA with a volume-equivalent diameter of $D_{\rm lit} = 46.9 \pm 0.3$~km from thermal measurement (WISE) \citep{2022PSJ.....3...56H}. This system was observed twice by stellar occultation in France, on February 16, 2025\footnote{\url{https://gaiamoons.imcce.fr/occ.php?p=31761}},occulting a 10.4 magnitude star along the northern coast of Mediterranean sea and February 26, 2025\footnote{\url{https://gaiamoons.imcce.fr/occ.php?p=31764}}, occulting a 12.9 magnitude star across Spain and France (see Fig~\ref{figA6}). Occultation circumstances are displayed in Table~\ref{tableA1} and light curves are shown in Figure~\ref{figA4}~(l-q). Dedicated observation campaigns were organised accordingly, involving 5 and 13 observers, respectively (see Table~\ref{tab:1127}). The February 26 campaign, in particular, presents three positive chords separated in the sky plane at the time of occultation referred in Figure~\ref{fig3} (Right). According to \cite{2024A&A...688A..50L}, the wobble period is $12.77 \pm 0.11$~h, close to the rotation period of the primary body (12.74 h, \citealt{2018Icar..309..297H}). As a result, the binary system would be in synchronous rotation. Given the elongated shape derived from the shape model, associated with concerns about the distance between chord 3 and chords 1 and 2, we can make the hypotheses that: 1. the elongated shape of the primary induces, during rotation, an offset between the photocentre and the centre of mass that is falsely interpreted as the signature of a companion; or, 2. the elongated shape of the model derived by \cite{2016A&A...587A..48D} is an merge of a primary and a secondary object, each smaller than the object derived in the literature, but whose dynamics introduce a bias in both the determination of the shape model and the determination of the system size. As demonstrated several times, large flat surfaces in convex models are unrealistic. They can “hide” large concavities or appear in the case of binary objects as seen for example for the DAMIT shape model of (4337) Arecibo \citep{2023A&A...675A..24D}. These two hypotheses are discussed in the following subsections. 

\subsubsection{Single body hypothesis}

To investigate the first hypothesis, we fit an ellipse as depicted in Figure~\ref{fig3} (left), the reduced $\chi^2_{\rm pdf}$ is 3.84. This fit yields a surface-equivalent diameter of $D_{\rm equiv} = 51.8 \pm 12.2$~km, the mean value being larger by 10\% ($3.3$~km) than the value derived from thermal measurements. As depicted by \cite{2020MNRAS.499.4570H} based on DAMIT models, surface-equivalent diameter retrieved by occultation are on on average 6\% larger than volume-equivalent diameter, the derived diameter value is then close to thermal measurements. To refine the analysis, we use the shape model derived by \cite{2016A&A...586A.108H} onto the plane of the sky using the available pole solution as displayed in Figure~\ref{figA11} (a). The underlying photometric data used to derive this model originate from the Lowell Observatory survey, and the shape and spin state were obtained using standard light-curve inversion techniques. The initial pole was subsequently refined as explained in \cite{2018Icar..309..297H}. While light-curve inversion models are known to provide only an approximate representation of the true shape, particularly for small-scale or non-convex features \citep{2001Icar..153...24K,2001Icar..153...37K}, the spin-axis orientation is generally more robust. For this particular case, no quality flag is associated with the shape model of (1127) Mimi, but the pole orientation has been updated by \cite{2018Icar..309..297H} using WISE thermal data giving $\alpha = 169.0$~° and $\delta = -68.8$~°, the uncertainty for these value are evaluated at 10°. Consequently, the pole solution is considered reliable for the purpose of projecting the model onto the sky plane, whereas the detailed shape should be regarded as an approximation and is therefore critically examined in the following analysis. 
\\

A least-squares approach was used to rotate and project the shape model within the uncertainty range onto the plane of the sky so as to best match the observed occultation chords. We use as size prior the size equivalent to the single ellipse derived $D_{\rm equiv}$. The fitting procedure was implemented using SORA \citep{2022MNRAS.511.1167G} and Astropy \citep{2013A&A...558A..33A}, together with a routine exploring the range of possible configurations and selecting the solution that minimises the cost function. The solution was subsequently refined by enforcing the pole constraints and uncertainties, leading to pole coordinates of $\alpha_{\rm fit}=172.7 \pm 2.5$°, $\delta_{fit}=-73.3 \pm 1.5$° (ICRF) and prime meridian phase $\phi_{fit} = 214 \pm 5$° at J2000. We note that this small change in the pole coordinates (angular distance of 5.8°) is expected to lead to minor changes in the shape, but a full shape inversion is beyond the scope of this study. This consideration do not impact our conclusions. The result of this projection using prior scale is shown in Figure~\ref{figA11}. We then compare these results with previous occultations of Mimi attempted on November 11, 2019 and July 21, 2023 from the archived IOTA Occult database\footnote{\url{http://lunar-occultations.com/iota/occult4.htm}} \citep{2020MNRAS.499.4570H}, selected for their ability to constrain the model, namely the number of positive chords and the confidence in the data and the event recorded on 2025-02-16. The events and the applied method are described in Appendix~\ref{app:1}.
\\

The constraints derived from the February 26, 2025 event do not provide a consistent match to past occultation data tested in this work, even when considering initial model errors (see Fig~\ref{figA11}). Figures~(b) and (d) highlight a region beneath the projected shape model that is not accounted by the current shape solution, while Figure~(c) shows a projection of the 3D shape model that intersects a negative chord. This argument further supports a critical assessment of the hypothesis that (1127) Mimi is a single elongated body.

\begin{table}[ht!]
    \caption{Ellipse parameters for stellar occultations by 1127 Mimi for each hypothesis.}
    \renewcommand{\arraystretch}{1.2}
    \begin{tabular}{l r}
    \hline\hline
        \multicolumn{2}{c}{Hypothesis - Single body} \\
        \hline\hline
        \multicolumn{2}{c}{Mimi} \\
        \hline
        Apparent semi-major axis & $a = ( 30.1 \pm 5.6)$~km\\
        Apparent oblateness & $\epsilon = 0.26 \pm 0.21 $\\
        Position angle & $PA = ( 8.3 \pm 23.9 )\,^\circ$\\
        Equivalent radius & $R_{\rm equiv} = ( 25.9 \pm 6.1 )$~km\\
        Right Ascension$^{(i)}$ & $+06^\mathrm{h} 58^\mathrm{m} 20.31091^\mathrm{s}$ (3.85)~mas \\
        Declination & $+21^\circ\,20'\,15.8216"$ (1.79)~mas \\
        \hline\hline
        \multicolumn{2}{c}{Hypothesis - Satellite detection} \\
        \hline\hline
        \multicolumn{2}{c}{Mimi.1 - Primary} \\
        \hline
        Apparent semi-major axis & $a = ( 25.2 \pm 4.1)$~km\\
        Apparent oblateness & $\epsilon = 0.20 \pm 0.15 $\\
        Position angle & $PA = ( 0.0 \pm 10.0 )\,^\circ$\\
        Equivalent radius & $R_{\rm equiv} = ( 22.5 \pm 4.1 )$~km\\
        Right Ascension$^{(ii)}$ & $+06^\mathrm{h} 58^\mathrm{m} 20.31139^\mathrm{s}$ (10.88)~mas \\
        Declination & $+21^\circ\,20'\,15.8201"$ (5.82)~mas \\
        \hline
        \multicolumn{2}{c}{Mimi.2 - Satellite} \\
        \hline
        Equivalent radius & $R_{\rm equiv} = 8.5^{+10.0}_{-1.1}$~km\\
        Right Ascension$^{(iii)}$ & $+06^\mathrm{h} 58^\mathrm{m} 20.30930^\mathrm{s}$ (17.31)~mas \\
        Declination & $+21^\circ\,20'\,15.8316"$ (12.43)~mas \\
        \hline
    \end{tabular}
    \label{table2}
    \tablefoot{
    Astrometric positions are given in geocentric ICRF at TCA 2025-02-26 19:29:50.70. Reference ephemeris \#JPL68 is used for prediction. The scale in the asteroid's orbit is 1~mas = 1.10~km.
    \tablefoottext{i}{Centre of ellipse in the projected sky plane in km are given with respect to prediction \#JPL68 (see Fig~\ref{fig3}), $(f,g) = (-8.0, 2.0)$~km.}
    \tablefoottext{ii}{$(f,g) = (-0.5,0.4)$~km.}
    \tablefoottext{iii}{$(f,g = (-33.2, 12.3)$~km.}
    }
    
\end{table}

\subsubsection{Binary system hypothesis}

The second hypothesis assumes the presence of a satellite around Mimi. We consider two objects: the primary, whose silhouette is approximated by an ellipsoid occulting Stations 1 and 2, and a satellite occulting Station 3 (Fig~\ref{fig3} , right). Table~\ref{table2} summarises the results of the fits. The associated $\chi^2_{\rm pdf}$ value is 1.21 for the primary and is not well defined for the satellite as it is constrained using a single positive chords (Station 3) using the other ones as negatives. Considering this, uncertainties are given at $1\sigma$ regarding the sizes of the bodies. The primary body have a surface-equivalent diameter of $D_{\rm prim} = 45.1 \pm 8.2$~km. The diameter of the satellite is $D_{\rm sat} = 17.0 \pm 2.2$~km and corresponds to a circle centred at the middle of the single occulted chord 3 with uncertainties driven by emersion and immersion times. It should be considered as a lower limit, the upper limit gives a value of 37~km. The derived total surface-equivalent radius (primary + satellite using the lower limit, of the system is $D_{\rm equiv} = 48.2 \pm 7.8$~km, consistent with literature value $D_{\rm lit}$. We emphasize that the size of the two objects is not well constrained by the observational data, so we cannot confidently discriminate between the two intervals of derived mass ratio. However, we note that for the lower limit value of the satellite, the associated mass-ratio is $0.0535 \pm 0.0351$, which is consistent with $q_1 = 0.019 \pm 0.001$ computed by \cite{2024A&A...688A..50L}, assuming that the derived diameters can stand for the volume-equivalent ones. 
\\

To further assess the binary hypothesis, we used the method described by \cite{2024A&A...688L..23L} to fit two possible orbit solutions for 1127 Mimi mutual system. To do so we used our position of the satellite derived by stellar occultation (Tab~\ref{table2}) and Gaia photometric data. The orbit parameters are given in Table~\ref{orbite}. For both orbit solutions, the implied mass ratio can be computed using the formula  $q=\,(m_{\rm sat}/m_{\rm prim}) \sim sf^{3/2}$, with f being the flux factor between the two bodies and $s$ a correcting factor to quantify the difference in composition. If $s$ differs significantly from unity, it suggests a mix of non-spherical shape, a different density, or a different albedo \citep{2024A&A...688L..23L}.
Solution 1 stands for a circular equatorial orbit. The associated mass ratio $q_{\rm sol1} = 0.0982$ (Table~\ref{orbite}) is consistent with occultation data, and leads to a satellite with a diameter of 19 km, considering the mean derived diameter of the primary. We note that by suggesting a scale factor to 1, according to the hypothesis of the same density and albedo used by \cite{2024A&A...688A..50L}, mass ratio becomes $0.021$, coherent with derived values from both occultation and Gaia data. However, the associated period (22.35~h) (Table~\ref{orbite}) is inconsistent with the wobble period value measured by Gaia (12.744~h \citealt{2024A&A...688A..50L}).
Solution 2 describes a slightly elliptical inclined orbit. Its associated mass ratio $q_{\rm sol2} = 0.1288$ gives a satellite 21~km wide, consistent with observations. The associated period (12.10~h) is close to the wobble period. Additionally, we propagate the position of the satellite at different times, as used in Appendix~\ref{app:1} - February 16, 2025, July 21, 2023 and November 19, 2019 - to confirm their viability. 
We note that solution 1 gives a position located at (60,5) km in Figure~\ref{figA11}~d, crossing a negative observation, considering the timescale of the propagation (6 years). This is not considered as critical since the satellite uncertainty position could be quite large. Apart from that observation, no clear inconsistencies were found as the satellite position are given outside the range observed by occultation. Taken together, the orbital elements and physical ratios inferred slightly favour the inclined orbit solution 2.
\\

The data currently available for 1127 Mimi do not allow us, at this stage, to conclude on the binary or non-binary nature of the object. The most effective way to discriminate between these two hypotheses is to acquire new stellar occultation data; indeed, a negative chord between chords 1 and 3 would have allowed us to make a conclusion on the binary nature.

\subsection{(36882) 2000 SW$_{155}$ on August 29, 2025}
\label{results:SW155}

The occultation by (36882) 2000 SW$_{155}$ (denoted SW155 in the following) on 29 August 2025\footnote{\url{https://gaiamoons.imcce.fr/occ.php?p=38307}} was observed from six stations across Europe. The shadow of this system was visible from 22:48:30 to 23:06:30 UT (see Fig~\ref{figA6}). The 9.3 km body (WISE, \citealt{2011ApJ...741...68M}) occulted a 11.2 magnitude star, the general circumstances of the occultation are given in Table~\ref{tableA1}. As the first recorded event for this object, two positive detections and three negative chords were recorded, as summarised in Table~\ref{tab:36882}. According to \cite{2024A&A...688A..50L}, the size of the possible satellite is expected to be between 0.6 and 4.6~km and a separation around 130~km maximum. 
\\

\begin{figure}[ht!]
\centering
\includegraphics[width=\hsize]{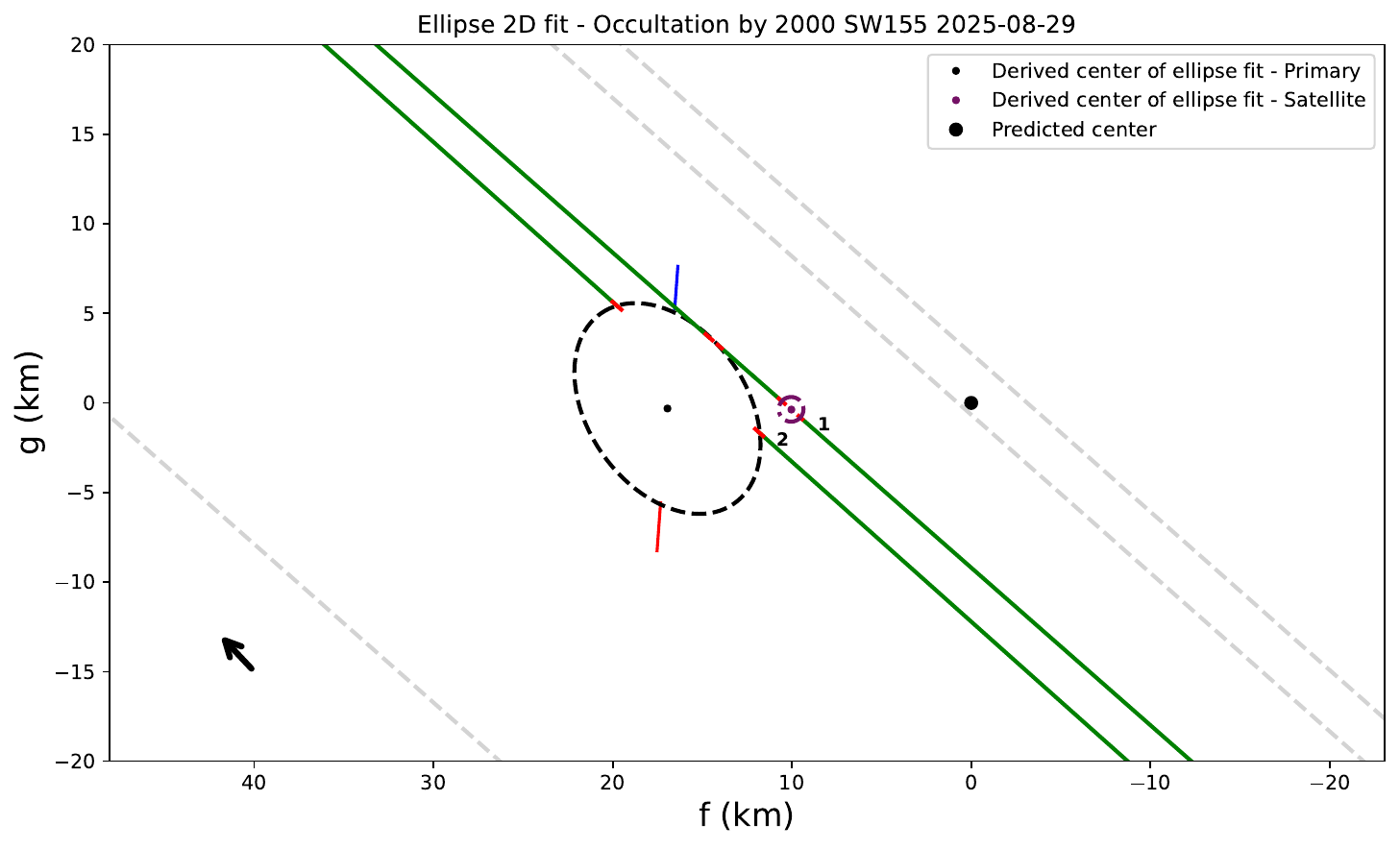}
  \caption{Sky plane projections of SW155 stellar occultation event of August 29, 2025 (see Table~\ref{tab:36882}). The lower estimation of the satellite is displayed in purple. Legends are the same as in Figure~\ref{fig2}. Fit details are given in Table\ref{tabled1}. The pole \citep{2023A&A...675A..24D} is represented by the red (north pole) and blue (south pole) axes to illustrate the proximity between the satellite and the equator of the primary. }
     \label{figd1}
\end{figure}

The estimated magnitude drop was $7.2 \pm 0.2$ ($\sim 99.87 \pm 0.02$\% flux drop), considering the apparent visible magnitude of the body ($18.4 \pm 0.2$, \citealt{2024A&A...687A..38C}). Photometric analysis of the positive light curves reveals a single positive drop of 0.84 s recorded at Station 2, and a double positive drop observed at Station 1 (see Fig~\ref{figA4}~(r-s)). The two drops are separated by 370~ms and last respectively for 120~ms and 90~ms with depths of about 80\% and 70\%, respectively. 
The hypothesis of a double star occulted by a single body is excluded as the double drop is not measured at Station 1 and the magnitude amplitude associated to this drop is significantly larger than 1. Consequently, the two drops measured at Station 2 correspond to two occulting bodies (see sky plane projection given Figure~\ref{figd1}). The first drop can be attributed to either a small kilometre-size object or the edge of another (or the same) object. As described by \cite{1987AJ.....93.1549R}, small circular objects (on the order of a few kilometres) and edge effects enhance diffraction effects on measured light curves, which tends to reduce the depth of the drops. Accounting for these effects (see Section~\ref{sec:method}), the minimum derived size for such a structure is 1.4~km if we approximate it as a circle, given that the corresponding Fresnel scale is 305~m. By convolving these effects with the exposure time 80~ms the light curve will be smoothed and will contribute to attenuate the light drop. The second drop corresponds to a highly grazing observation. As diffraction effects depend on the normal local surface velocity they are particularly significant in this case and explain the attenuation and overestimation of the associated drop duration. The configuration shown in Figure~\ref{figd1} allowed us to estimate an attack angle of about 75°. The separation between the two drops recorded at Station 2 corresponds to a projected distance of 4~km in the plane of the sky at the asteroid’s geocentric distance (1.9~AU). We note that the two negative chords lie close to the central line of the predicted position of the primary. A significant offset of the system barycentre was observed. This is consistent with the ephemeris uncertainties, namely 17~mas in right ascension and 3~mas in declination ($3\sigma$). The resulting silhouette therefore exhibits ambiguous characteristics (Fig~\ref{figd1}). One possible interpretation is that of a single object presenting a local concavity or limb feature. Although such a configuration could in principle reproduce the observed light curve, it appears unlikely in view of the overall consistency of the occultation chords. This hypothesis is therefore considered less probable, though it cannot be completely excluded. 
\\

We therefored consider an alternative interpretation in which the second event is caused by a satellite orbiting the primary body. The derived parameters of the system, together with the relative positions of the components in the plane of the sky at TCA, are summarised in Table~\ref{tabled1}. This fit yields a minimum estimated satellite radius of 690~m, the maximum value being 3.2~km considering the limits imposed by negative chords 5 and 6. The primary body has an elongated shape with semi-major radius of $6.4 \pm 1.1$~km and an oblateness of $0.27 \pm 0.10$ (arbitrary value). The two bodies are displayed in Figure~\ref{figd1}. Taken together, the total surface-equivalent diameter of the primary and the satellite are set between $D_{\rm min} = 10.2 $~km and $D_{\rm max} = 13.6$~km and the associated mass ratios are $q_{\rm min} = 0.0021$ and $q_{\rm max} = 0.1970$. The lower limit value of the surface-equivalent diameter is consistent with WISE thermal measurement. The shape of the primary is chosen elongated as suggested by Gaia photometric observations. This prior is consistent with the present stellar occultation observation as well, though it is not considered as critical for the binary hypothesis assessment. Finally, this allowed us to discriminate the two values of mass-ratio estimated by \cite{2024A&A...688A..50L} and \cite{2026arXiv260522702L}, $q_1 = 0.0032 \pm 0.0022$ and $q_2 = 0.884 \pm 0.115$, the lower value of mass-ratio obtained by our fit and the lower estimation $q_1$ are consistent and is a valuable argument towards the satellite hypothesis. This way, 2000 SW$_{155}$ stands as a strong binary candidate. Further observations will be necessary to fully describe this system.

\begin{table}[ht!]
    \caption{Ellipse parameters for stellar occultations by SW155 for each hypothesis.}
    \renewcommand{\arraystretch}{1.2}
    \begin{tabular}{l r}
        \hline\hline
        \multicolumn{2}{c}{SW155.1 - Primary} \\
        \hline
        Apparent semi-major axis & $a = ( 6.4 \pm 1.1)$~km\\
        Apparent oblateness & $\epsilon = 0.27 \pm 0.10 $\\
        Position angle & $PA = ( 125.0 \pm 20.0 )$~deg\\
        Equivalent radius & $R_{\rm equiv} = ( 5.5 \pm 1.0 )$~km\\
        Right Ascension$^{(i)}$ & $+20^\mathrm{h} 53^\mathrm{m} 5.89451^\mathrm{s}$ (0.57)~mas \\
        Declination &$-00^\circ\,33'\,58.0994"$ (0.85)~mas \\
        \hline
        \multicolumn{2}{c}{SW155.2 - Satellite} \\
        \hline
        Equivalent radius & $R_{\rm equiv} = (0.7 \pm 0.3)$~km\\
        Right Ascension$^{(ii)}$ & $+20^\mathrm{h} 53^\mathrm{m} 5.89416^\mathrm{s}$ (0.21)~mas \\
        Declination & $-00^\circ\,33'\,58.0993"$ (0.25)~mas \\
        \hline
    \end{tabular}
    \label{tabled1}
    \tablefoot{
    Astrometric positions are given in geocentric ICRF at TCA 2025-08-29 22:57:30.74. Ephemeris are given with respect to the prediction JPL\#45. The scale in the asteroid's orbit is 1~mas = 1.38~km.
    \tablefoottext{i}{centre of ellipse in the projected sky plane in km (Fig~\ref{figd1}) $(f,g) = (17.7, -0.6)$~km.}
    \tablefoottext{ii}{$(f,g) = (10.4,-0.5)$~km. The solution displayed for the satellite corresponds to a lower limit associated to the centre of the first drop measured by Station 3.}
    }
    
\end{table}

\section{Continuous follow-up of potential binary systems}
\label{sec:summary}

We present here a summary of occultation campaigns with events providing relevant new information for a given asteroidal system - namely at least two positive chords or a sufficient number of negative chords probing the surrounding environment. Identification information for the system and the characteristics of the event (date, number of chords), as well as the surface-equivalent radius compared with the most relevant literature value \citep{2023A&A...671A.151B} and the derived astrometric position are given for 28 events over 20 sytems in Table~\ref{tableCampaign}. These events were recorded from October 2023 to February 2026, excluding those already discussed above. The corresponding ellipsoidal fits are available in Figure~\ref{figA51}. When available and relevant (i.e. more constraining and better fitting), the 3D model is used to derive the diameter and the astrometric position,  the results are given in the fifth column of Table~\ref{tableCampaign}. Some events exhibit key details that warrant further examination. We note that the shape model available for (370) Modestia is not compatible with the occultation event on December 30, 2025. An elliptical fit was performed and revealed an elongated shape (Figure~\ref{figA51}). This consideration raises the question of a potential close configuration if this object has a satellite. Similarly, (1109) Tata observation on February 27 highlights an elliptical object with a surface diameter significantly larger than literature values. (550) Senta has been observed ten times by occultations since 2024, with nine observations with at least one positive chord. The three events presented in Figure~\ref{figA51} add more value for the shape model (one model can be discarded thanks to the September 7, 2025 event). No observations of the satellite has been recorded as the direct environment of this object has not been investigated enough. The event of September 25, 2025 constitutes the largest occultation observation for (712) Boliviana, no satellites has been observed but critical values in term of size and astrometry are derived. Finally, (4337) Arecibo has been observed twice with its satellite in the United States, adding positions to model its mutual orbit with accuracy and study its characteristics.

\begin{table*}[ht!]
\begin{center}
    \small
    \caption{Number of chords, derived surface diameter, and astrometric coordinates of the most successful stellar occultation campaigns organised within the GaiaMoons project framework.}
    \renewcommand{\arraystretch}{1.2}
    \begin{tabular}{l c c c c c r r}
        \hline
        ID & Date at TCA (UT) & ($N_T,\, N_P,\, N_N)^{(i)}$ & $R_{\rm equiv}$ & $R_{ref}^{(ii)}$ & Right Ascension  & Declination \\
        & & & km (km)& km (km) & hh mm ss.sss (mas)  & dd mm ss.sss (mas) \\
        \hline
        31 & 2024-07-01 09:33:39 & (2,2,0) & 134.1 (8.2)  & 134 (2.3) & $+11^\mathrm{h} 08^\mathrm{m} 11.11492^\mathrm{s}$ (3.03) & $+19^\circ\,55'\,28.5729"$ (2.10)\\
        53 & 2025-08-17 02:42:15 & (4,2,2) & 51.4 (5.5) & 51.4 (0.9) & $+21^\mathrm{h} 16^\mathrm{m} 31.94997^\mathrm{s}$ (5.13) & $-15^\circ\,14'\,51.9664"$ (10.18)\\
        146 & 2025-11-13 17:56:18 & (4,4,0) & 63.6 (0.8) & 65.5 (7.5) & $+07^\mathrm{h} 12^\mathrm{m}$ $51.24139^\mathrm{s}$ (1.23)& $+24^\circ\,14'\,24.8121"$ (0.67)\\
        146 & 2025-12-18 01:07:24 & (23,3,7) & 65.5 (6.5) & 65.5 (7.5) & $+06^\mathrm{h} 55^\mathrm{m} 17.71884^\mathrm{s}$ (7.24) & $+27^\circ\,17'\,48.0841"$ (7.46)\\
        206 & 2025-12-19 01:19:18 & (2,2,0) & 47.0 (2.4) & 45.8 (1.5) & $+07^\mathrm{h} 09^\mathrm{m} 47.69612^\mathrm{s}$ (2.28) & $+18^\circ\,13'\,09.1364"$ (4.09)\\
        247 & 2025-01-01 18:21:42 & (2,2,0) & 75.3 (3.8) & 75.3 (1.6)$^*$ & $+05^\mathrm{h} 53^\mathrm{m} 50.57195^\mathrm{s}$ (6.48) & $+66^\circ\,01'\,21.7315"$ (6.26)\\
        247 & 2025-03-02 22:23:56 & (6,4,2) & 73.0 (4.4) & 75.3 (1.6)$^*$  & $+05^\mathrm{h} 56^\mathrm{m} 59.46409^\mathrm{s}$ (2.28) & $+52^\circ\,25'\,36.7209"$ (3.04)\\
        264 & 2024-06-08 04:55:57 & (2,2,0) & 26.3 (2.6) & 26.4 (1.1) & $+10^\mathrm{h} 58^\mathrm{m} 22.39671^\mathrm{s}$ (0.91) & $+15^\circ\,25'\,29.7571"$ (0.92)\\
        370 & 2025-12-30 10:16:24 & (4,3,1) & 20.3 (1.5) & 19.1 (0.1) & $+06^\mathrm{h} 18^\mathrm{m} 00.45154^\mathrm{s}$ (0.68) & $+26^\circ\,53'\,51.5843"$ (0.95)\\
        542 & 2025-02-05 00:08:57 & (10,3,6) & 23.6 (1.6) & 24.2 (0.2) & $+07^\mathrm{h} 41^\mathrm{m} 08.35231^\mathrm{s}$ (1.11) & $+12^\circ\,08'\,01.9050"$ (1.49)\\
        550 & 2024-03-21 04:29:31 & (2,2,0) & 18.9 (1.9) & 18.9 (0.5) & $+18^\mathrm{h} 41^\mathrm{m} 42.76069^\mathrm{s}$ (1.28) &  $-26^\circ\,16'\,50.5429"$ (2.56)\\
        550 & 2024-07-01 09:33:39 & (6,5,1) & 19.1 (0.1) & 18.9 (0.5)  & $+19^\mathrm{h} 38^\mathrm{m} 54.99258^\mathrm{s}$ (0.27) & $-10^\circ\,58'\,12.3179"$ (0.44)\\
        550 & 2025-09-07 00:54:18 & (4,2,0) & 19.0 (1.8) & 18.9 (0.5)  & $+05^\mathrm{h} 49^\mathrm{m} 21.35555^\mathrm{s}$ (2.55) & $+27^\circ\,36'\,37.5996"$ (7.69)\\
        712 & 2025-07-26 01:44:55 & (11,1,9) & 59.8 (12.3) & 59.8 (2.0)$^*$ & $+20^\mathrm{h} 59^\mathrm{m} 42.01736^\mathrm{s}$ (16.43) & $+04^\circ\,10'\,46.5957"$ (42.68)\\
        712 & 2025-09-25 22:26:50 & (33, 15, 11) & 55.2 (7.2) & 59.8 (2.0)$^*$ & $+20^\mathrm{h} 23^\mathrm{m} 36.51102^\mathrm{s}$ (5.28)  & $-00^\circ\,57'\,40.6986"$ (3.35)\\
        1024 & 2025-11-05 20:48:55 & (20,3,11) & 19.7 (0.9) & 20.9 (0.9) & $+02^\mathrm{h} 10^\mathrm{m} 52.06946^\mathrm{s}$ (1.05) & $+01^\circ\,36'\,16.0688"$ (1.06)\\
        1800 & 2025-04-04 20:30:47 & (7,2,1) & 3.6 (0.6) & 3.6 (0.1) & $+12^\mathrm{h} 54^\mathrm{m} 06.50078^\mathrm{s}$ (0.88) & $+03^\circ\,39'\,52.4196"$ (1.34) \\
        2388 & 2025-11-17 22:24:58 & (9,3,4) & 4.5 (0.9) & 4.1 (0.9) & $+05^\mathrm{h} 28^\mathrm{m} 04.46662^\mathrm{s}$ (1.15) & $+26^\circ\,22'\,08.6864"$ (0.75)\\
        3728 & 2025-06-16 23:40:51 & (12,2,9) & 11.8 (1.0) & 12.1 (0.1) & $+19^\mathrm{h} 39^\mathrm{m} 10.74449^\mathrm{s}$ (0.54) & $+06^\circ\,17'\,24.2765"$ (1.89)\\
        4332 &  2025-11-29 14:41:07 & (12,3,9) & 5.4 (2.1) & 5.4 (0.9) & $+07^\mathrm{h} 20^\mathrm{m} 46.46990^\mathrm{s}$ (2.69) & $-05^\circ\,40'\,58.5082"$ (2.53)\\
        4337a & 2024-09-14 09:17:23 & (8,4,4) & 8.9 (0.9) & 12.2 (0.3)$^{**}$ & $+07^\mathrm{h} 00^\mathrm{m} 15.69369^\mathrm{s}$ (0.94) & $+24^\circ\,13'\,21.4692"$ (1.22)\\
        4337b & 2024-09-14 09:17:23 & (8,4,4) & 5.4 (1.4) & 6.5 (0.7)$^{**}$ & $+07^\mathrm{h} 00^\mathrm{m} 15.69345^\mathrm{s}$ (0.53) & $+24^\circ\,13'\,21.4730"$ (0.71)\\
        4337a &  2024-10-16 08:41:28 & (5,4,1) & 9.1 (1.5) & 12.2 (0.3)$^{**}$ & $+07^\mathrm{h} 25^\mathrm{m} 44.69838^\mathrm{s}$ (0.44) & $+23^\circ\,49'\,40.1676"$ (0.77)\\
        4337b &  2024-10-16 08:41:28 & (5,4,1) & 5.6 (0.9) & 6.5 (0.7)$^{**}$ & $+07^\mathrm{h} 25^\mathrm{m} 44.69979^\mathrm{s}$ (1.54) & $+23^\circ\,49'\,40.1629"$ (0.46)\\
        4999 & 2025-09-16 05:58:52 & (8,4,4) & 5.9 (1.4) & 7.2 (0.6) & $+18^\mathrm{h} 30^\mathrm{m} 7.73005^\mathrm{s}$ (0.33) & $-15^\circ\,26'\,29.0989"$ (0.65)\\
        13147 & 2025-12-04 20:46:57 & (2,2,0) & 3.2 (0.4) & 3.2 (0.2) & $+13^\mathrm{h} 20^\mathrm{m} 47.57791^\mathrm{s}$ (0.73) & $-06^\circ\,38'\,58.3287"$ (0.75) \\
        2731 & 2026-02-09 21:57:06 & (3,2,0) & 24.6 (1.3) & 22.6 (1.1) & $+08^\mathrm{h} 11^\mathrm{m} 45.35583^\mathrm{s}$ (0.31) & $+12^\circ\,48'\,56.2811"$ (0.50) \\
        1109 & 2026-02-27 22:18:00 & (12,4,4) & 36.9 (3.0) & 32.0 (1.3) & $+08^\mathrm{h} 00^\mathrm{m} 24.96611^\mathrm{s}$ (0.17) &  $+16^\circ\,54'\,31.9930"$ (1.03) \\
        \hline
    \end{tabular}
    \tablefoot{Astrometric positions are given in geocentric ICRF at the given date. Derived uncertainties correspond to $3\sigma$ values, literature values are given in $1\sigma$ uncertainties, unless specified. Suffix a and b are respectively given for the primary and the satellite of (4337) Arecibo known binary system.
    \tablefoottext{i}{$N_T$ total number of stations, $N_P$ number of positive stations, $N_N$ number of negative stations, remaining stations were overcast or had technical issues.}
    \tablefoottext{ii}{Reference values are retrieved from SsODNet database \citep{2023A&A...671A.151B}. Reference values are given as volume-equivalent radii, values given as surface-equivalent radii are marked with an upper asterisk ($^*$).}
    \tablefoottext{**}{Values are retrieved from occultation by \cite{2022MPBu...49....3G}, thoses values were performed using a circular fit, the present study used an elliptical fit (see Fig~\ref{figA51}}).
    For the following systems, information have been computed using their shape model (model on DAMIT): Euphrosyne (4399), Lucina (1837), Eukrate (1207), Senta (774). Among the two pole models available in the DAMIT database for 550 Senta, model 774 was chosen as it provided a better fit to the observational data.}
    \label{tableCampaign}
\end{center}
\end{table*}

\section{Discussion}
\label{sec:improve}

\begin{figure}[ht!]
\centering
    \includegraphics[width=\hsize]{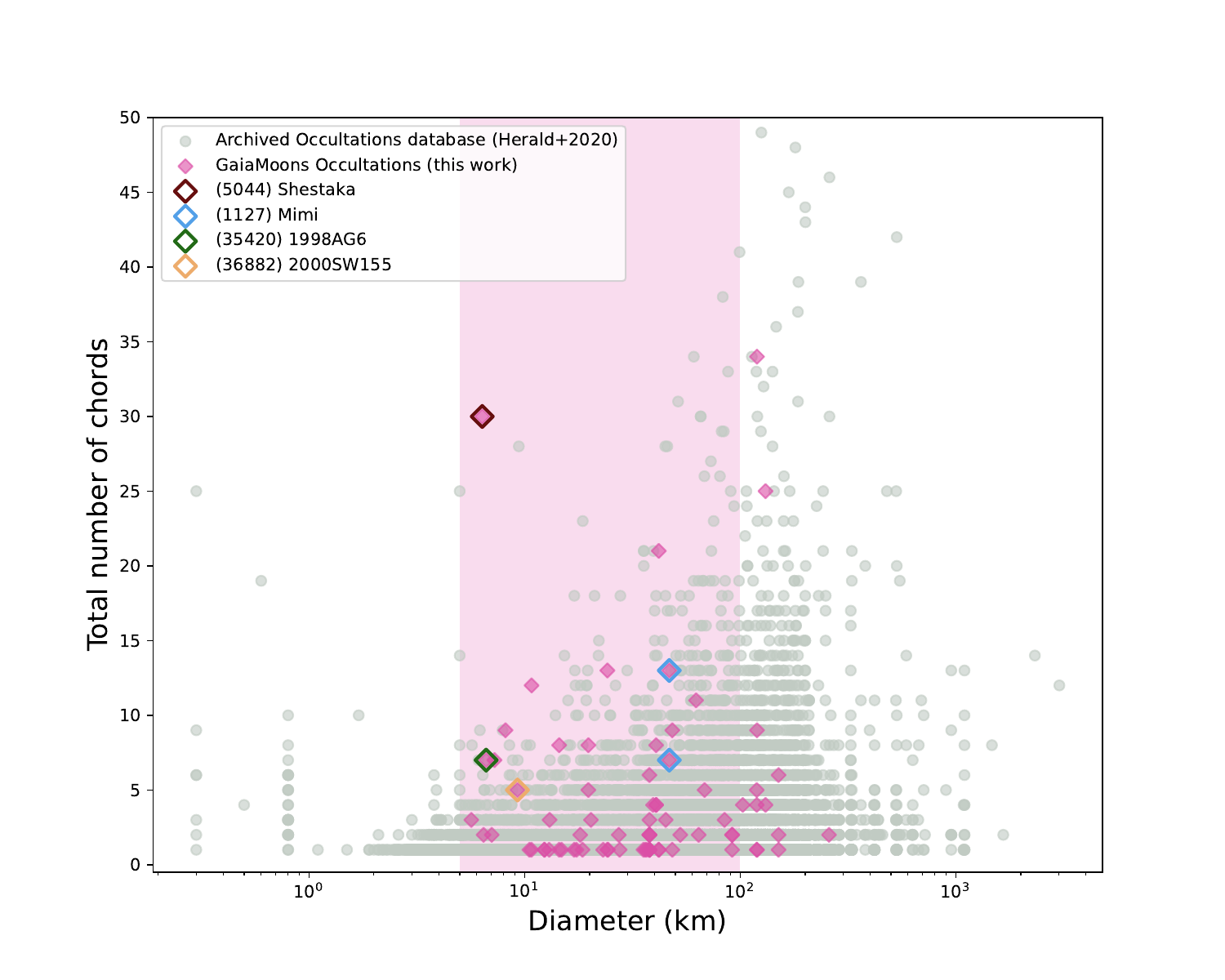}
    \caption{Distribution of the total number of chords for each occultation event associated with the target diameter. Events observed within the framework of GaiaMoons (October 2023 - February 2026) are displayed in purple diamond, and main events presented in this paper are highlighted with coloured contours. Archived occultations were retrieved from the IOTA Occult database, compiling observations for a period between 1961 and 2025 for small bodies of the Solar System \citep{2020MNRAS.499.4570H}, and they are displayed in grey. The purple area in the background represents the focus space zone targeted by the GaiaMoons program.}
    \label{fig-distrib}
\end{figure}

 While stellar occultations have so far primarily yielded dense chord coverage for large objects ($D > 100\,\mathrm{km}$), the present study deliberately targets intermediate-size systems in order to maximise the number of positive chords in a poorly explored size regime. This domain, shown as the purple focus space region in Figure~\ref{fig-distrib}, represents a key parameter space where additional occultation measurements can significantly improve the characterisation of binary asteroid systems. Among the targets that we observed within the framework of the GaiaMoons project and for which we obtained at least one positive chord, half have a diameter of less than 38~km, and 90\% of them have a diameter of less than 100~km.

Stellar occultations constitutes a robust and powerful technique to characterise asteroid systems, they make it easier to distinguish “distant” binary asteroids from contact binary asteroids compared to traditional methods, such as photometry for double synchronous systems. For occultations, this ability of disentanglement is directly correlated to the number of observers and also depends on the geometry and the size of the system at the moment of the occultation. Some examples have been described in Sections \ref{results:AG6} and \ref{results:mimi} for AG6 and Mimi. The value of the astrometric constrain also depends on the number of observers, as described by \cite{2024A&ARv..32....6S}, for intermediate-size objects, the number of stations to reach a kilometric precision in its position is about 20 with around seven positives among them, (5044) Shestaka (Section \ref{results:shestaka}) constitutes a typical application of this consideration. Of course, the number of observer will depend as well on the networks availability and position around the globe, as this study rely on the communities known by the scientific team, there is a bias in the occultation observation location mostly over Europe, USA and East Asia.

Considering the tools used for analysis, since SORA was designed for occultations of relatively large objects (TNOs, Trojans, \citealt{2022MNRAS.511.1167G}), diffraction effects are usually considered negligible and as a result are not adequately considered (see Sections \ref{results:AG6} and  \ref{results:SW155}) for intermediate objects in certain particular orientations. For binary systems, a more faithful modelisation would be to apply the model described by \cite{1987AJ.....93.1549R} to quantitatively compute the Fresnel fringes of multiple bodies with ellipsoidal shapes, which could be done in a future work and become applicable for stellar occultation observations by small MBAs and NEOs.
\\

Overall, these observation campaigns allow access to a precise shape with kilometre accuracy of the occulted objects and a discussion of their origins. We further note that the inferred location of the potential satellite of SW155 is close to the equatorial plane of the primary. Figure~\ref{figd1} shows a projection of the spin pole \citep{2023A&A...675A..24D} in the plane of the sky, with the satellite position indicated for illustration. For small objects ($D<10$~km), such as AG6 and SW155, satellite formation scenarios involving rotational disruption driven by YORP\footnote{Yarkovsky–O’Keefe–Radzievskii–Paddack}-induced spin-up, which can lead to equatorial mass shedding and subsequent re-accumulation into a secondary component \citep{2007ApJ...659L..57C,2008Natur.454..188W} are possible.  In this case, the limited size of the satellite, estimated to be approximately 10\% of the primary’s size, makes its detection particularly challenging using classical photometric methods, where its signal can be easily masked by shape effects or rotational variability. Similarly, radar observations are not feasible for this object given its distance and size. Additionally we note that AG6 suggests a bi-lobed shape, such contact binary asteroids can also form through an uneven mass distribution combined with YORP spin-up \citep{2021MNRAS.507.4914Z,2011Icar..214..161J}.
Alternatively, this shape is similar to that of asteroid (216) Kleopatra. \cite{2021A&A...653A..57M} discuss the origin of such a shape, which could also be due to a large impact followed by a re-accumulation resulting in two distinct merged lobes. Both objects have close orbital parameters as well \citep{2024ApJS..274...25N}. However, the difference in size between the two objects (118 km versus 6.6 km) puts the viability of such a scenario into perspective. YORP-induced mechanism would also explain the origin of the potential satellite of 1127 Mimi, but would be unlikely due to its larger size. The formation mechanism would be the similar 4337 Arecibo. The results are compatible with Escaping Ejecta Binaries \citep{2004Icar..170..243D} which is a currently under-represented population of binary systems formed through sub-catastrophic collisions

\section{Conclusions and perspectives}
\label{sec:ccl}

Through this work we have presented new physical and astrometric constraints derived from stellar occultations for 101 target asteroid systems observed between October 2023 and February 2026. This study represents the first dedicated effort targeting intermediate-size objects in the 5–100 km range through a coordinated occultation strategy, rather than isolated single-object campaigns. In particular, the occultation of (5044) Shestaka on October 23, 2024, gathered an unprecedented number of participants for an object of this size. Four detailed campaigns were presented as case studies, highlighting both the challenges inherent to binary asteroid occultation observations and the unique datasets that can be obtained in terms of shape, size, and astrometry. It should be noted that some satellites may have gone undetected due to noise, diffraction, or grazing effects observed in certain light curves as well as differences in albedo between the two bodies. To date, (36882) 2000 SW$_{155}$ exhibits strong indications of binarity. (35420) 1998 AG$_6$ and (206) Hersilia have strong binary or contact-binary interpretations, and (1127) Mimi remains a slightly more ambiguous case. These targets will remain the focus of follow-up stellar occultation campaigns in the future. Seventy-six observations led to at least one positive chord, and of these 33 allowed valuable new information to constrain the size and the position of the targets. This number includes the stellar occultation events presented in Section~\ref{sec:results} and Section~\ref{sec:summary}. All occultation measurements obtained in this work have been made publicly available, enabling continued follow-up and independent confirmation of candidate systems. 

To date, many satellite detections through occultations remain largely serendipitous. A major objective of current projects is therefore to extend these discoveries into a more systematic and reproducible methodology. This strategy naturally complements established large-scale occultation programs, such as Lucky Star, that have successfully characterised TNOs and Jupiter Trojans. Similar approaches are also envisioned for cometary nuclei in an attempt to constrain their ephemerides and shapes \citep{2025RSPTA.38340189P,2018JOA.....8a..11M}. On a broader scale, dedicated efforts, such as those gathered around the GaiaMoons program, have organised dozens of coordinated campaigns, demonstrating both the feasibility and the logistical challenges of systematically exploiting occultations with Gaia data. Our study and the Gaiamoons project demonstrate the efficiency of mixing highly precise Gaia data with stellar occultation to characterise new binary or contact binary systems.  Each successful observation adds a constraint of the shape and the position of the components, refines orbital predictions, and increases the efficiency of subsequent campaigns. By applying a reproducible methodology, GaiaMoons provides critical constraints in a poorly explored regime of parameter space and helps resolve ambiguities left by other observational techniques. In particular, stellar occultation can distinguish a wide range of mass and size ratios (from nearly equal-mass binaries to small satellites), and this technique offers essential inputs to refine approximate 3D shape reconstruction models, for instance, through ADAM-based approaches \citep{2015A&A...576A...8V}. Considering the new data from Gaia that will be released in the coming years, the method presented in this work is a promising way to study binary systems in the coming decades.
\\

Looking ahead, the forthcoming Extremely Large Telescope (ELT) will dramatically transform the landscape of small-body studies. The ELT will open additional observational opportunities through direct imaging. With an expected angular resolution of about 3 mas in the visible wavelength ($\lambda \sim 0.5$~$\mu$m), the ELT may enable the direct resolution of a new population of intermediate-size binary systems in the main belt, thus adding a potential new method to characterise intermediate-size binaries.

\section{Data availability}
The data used in this study, including light curves, derived immersion and emersion timings and stations information, are only available in electronic format the CDS via anonymous ftp to \url{cdsarc.u-strasbg.fr} (130.79.128.5) or via \url{http://cdsweb.u-strasbg.fr/cgi-bin/qcat?J/A+A/.}. Additional data not included in this article but part of the overall Gaiamoons occultation observation dataset will be available in the thesis manuscript of the corresponding author that will be published in late 2026. aw data of this study are available from the corresponding author upon request.

\begin{acknowledgements}
This work was supported by the project GaiaMoons of the Agence Nationale de Recherche (France), grant ANR-22-CE490002. The GaiaMoons team gratefully acknowledges the amateur communities of IOTA, IOTA/ES, IOTA/EA, TTOA, and Planoccult for their essential support, dedication, and significant contributions to this work. The properties of Solar System Object are from the service SsODNet.ssoCard of the Space Service (SE-OP) of Laboratoire Temps Espace at Paris Observatory through its Solar System Portal (\url{https://ssp.imcce.fr}) \citep{2023A&A...671A.151B} with the python library rocks (\url{https://github.com/maxmahlke/rocks}). We made use of Astropy, a community-developed core Python package for Astronomy \citep{2013A&A...558A..33A}. This work has made use of data from the European Space Agency (ESA) mission Gaia (\url{https://www.cosmos.esa.int/Gaia}), processed by the Gaia Data Processing and Analysis Consortium (DPAC, \url{https://www. cosmos.esa.int/web/Gaia/dpac/consortium}). The authors would like to thank the Action Pluriannuelle Incitative (API) Pro-Am initiated and supported by Paris Observatory in the ROADIES program context. Z. Liu and D. Hestroffer thank the Academie Spatial program for their support. Part of observations were funded by the Scientific Research Projects Coordination Unit of Istanbul University with project numbers: BAP-3685 and FBG-2017-23943. Felipe Braga-Ribas acknowledges CNPq grant 316604/2023-2 and the financial support of the NAPI “Fenômenos Extremos do Universo” of Fundação de Apoio à Ciência, Tecnologia e Inovação do Paraná. M.A. thanks grants CNPq 427700/2018-3, 310683/2017-3, and 473002/2013-2, and FAPERJ E-26/210.705/2024. Y. Kilic, J.L. Ortiz, N. Morales and P. Santos-Sanz acknowledge financial support from the Severo Ochoa grant CEX2021-001131-S funded by MICIU/AEI/10.13039/501100011033. P. Santos-Sanz and Y. Kilic acknowledge financial support from the Spanish I+D+i project PID2022-139555NB-I00 (TNO-JWST) funded by MCIN/AEI/10.13039/501100011033. TUG100 Telescope at the Antalya TUG Site of the Türkiye National Observatories has been utilised, and we express our gratitude for the support provided by the Türkiye National Observatories and all its staff. The authors would like to thank the following observers who participated and provided data for the events: R. Leiva, A. Ossola, A. Catapano, B. Lade, R. Prentice, S. Whitehurst, E. Goni, E. Namur Neto, J. Delincak, J. Talbot, J. A. Berlanga, J. Collada Barcena, K. Harrison, M. Skrutskie, M. Gutekunst, N. Karaman, N. Wakefield, V. Cucchiaro, X. Dupont, T. Legault, S. Kidd, P. Le Cam, J.-N. Ferrier, G. Vanwalleghem, E. Vauthrin, F. Thill, B. Lott, D. Tamonis, and A. Maury.
\end{acknowledgements}

\bibliographystyle{bibtex/aa}  
\bibliography{main} 

\begin{appendix}
\onecolumn
\section{(1127) Mimi shape model analysis}
\label{app:1}

\begin{figure*}[ht!]
\centering
\includegraphics[width=.9\textwidth]{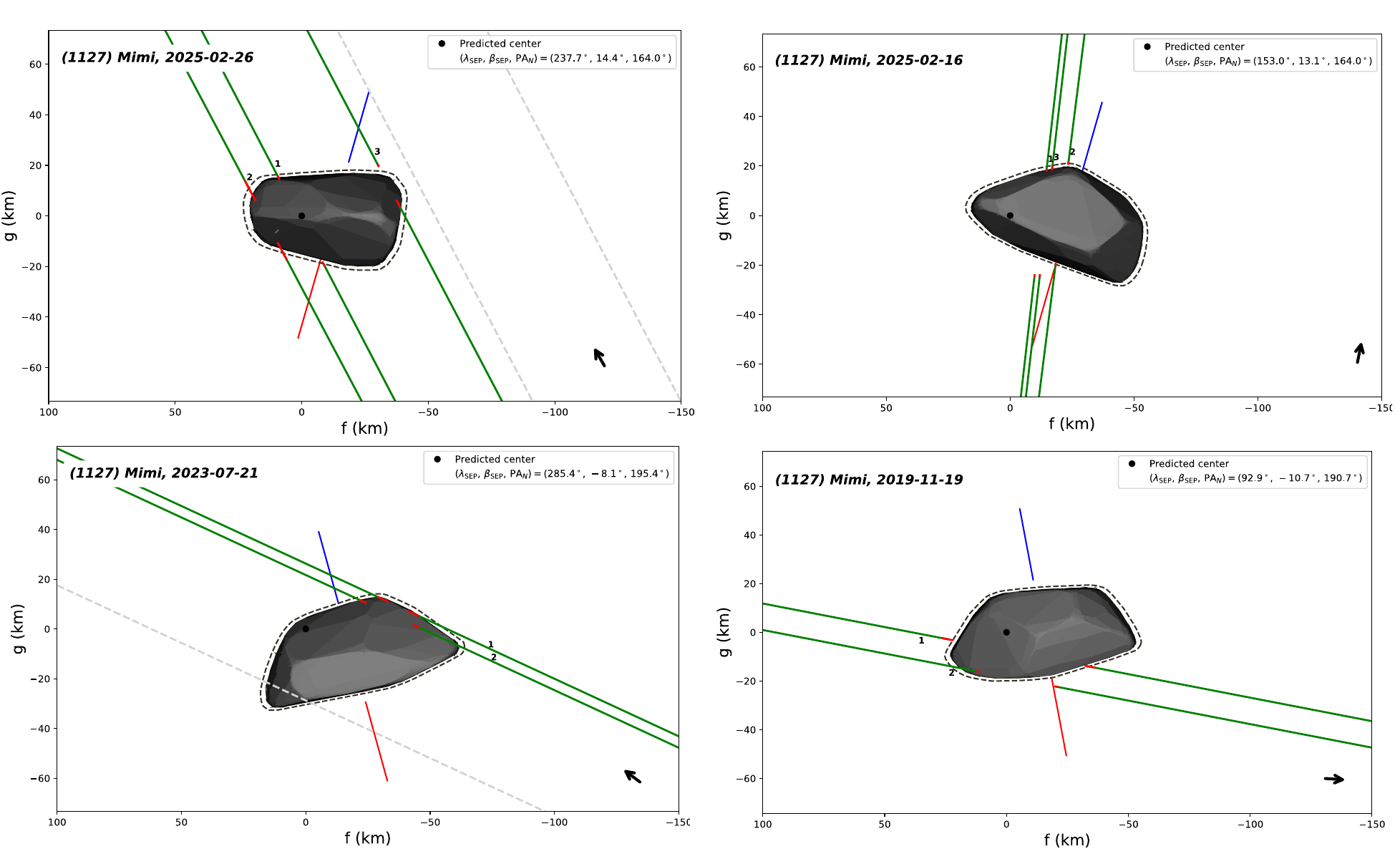}
  \caption{Adjustment of the shape model of (1127) Mimi using four stellar occultation events. Legends are the same as Figure~\ref{fig2}. The model was constrained using the data from event (a); events (b), (c), and (d) show the shape model after extrapolation of the constrained rotation model, with the centre shifted to best match the stellar occultation data.
  (a) $(f,g) = (-12.6,-0.5)$~km, TCA = 2025-02-26 19:29:50.700, 
  (b) $(f,g) = (-23,-3.5)$~km, TCA = 2025-02-16 01:28:56.240,
  (c) $(f,g) = (-19.0,-11.0)$~km, TCA = 2023-07-21 14:00:21.420,
  (d) $(f,g) = (-15.0,0.0)$~km, TCA = 2019-11-19 10:10:25.540.}
     \label{figA11}
\end{figure*}

The shape model orientation is constrained using occultation data from February 26, 2025. For past events, immersion and emersion times were retrieved from the Occult database \citep{2020MNRAS.499.4570H} and occultation profiles were reconstructed using SORA. Assuming a uniform rotation around the spin axis, given the object’s rotation period $P = 12.745$~h, the shape model is projected onto the local occultation sky plane by varying the position of the centre in order to achieve the best possible fit with the data. We start from an arbitrarily defined prior point, testing the different limits constrained by the literature, and vary the position of the centre accordingly. For each observation, resulting sub-observer coordinates $\lambda_{SEP}$ and $\beta_{SEP}$ as well as the pole position angle $PA_N$ are displayed in the top left legend section of the image in Figure~\ref{figA11}. We note that the uncertainty of the period here is for the most pessimistic estimate at $3\times 10^{-3}$~h \citep{2022FrASS...909771D}, it would result to a variation of the rotation by $\pm 17^\circ$ after 6 years, which does not significantly impact our conclusions.

\begin{table}[ht!]
\renewcommand{\arraystretch}{1.15}
\small
    \begin{center}
    \caption{Mimi mutual orbit parameters - best-fit solutions}
    \begin{tabular}{l r r}
        \hline\hline
         & Solution 1 & Solution 2 \\
        \hline
        Semi-major axis (a) & 66.9~km & 68.4~km \\
        Longitude of the Ascending Node ($\Omega$)  & 363.5° & 309.5° \\
        Inclination (i) & 05.8° & 30.5° \\
        Argument of pericentre ($\omega$) & -90.8° & -39.1°\\
        Eccentricity (e)  & 0.005 & 0.104 \\
        Pericentre relative time ($t_{\mathrm{peri}}^{(i)}$) & 0.93 day & 0.46 day \\
        Period (P) & 22.35 h & 12.10 h\\
        Scale parameter (s) & 4.9 & 1.9 \\
        Flux ratio (f) & 0.0738 & 0.1663 \\
        Mass ratio (q) & 0.0982 & 0.1288 \\
        \hline
        
    \end{tabular}
    \label{orbite}
    \end{center}
    \tablefoot{Orbital parameters for the system of 1127 Mimi, obtained using Gaia photometric data and the position of the primary and the satellite on February 26, 2025 at TCA.
    \tablefoottext{i}{Pericentre past time is computed as $t_{\mathrm{peri}} = t_0 + t_{\mathrm{ref}}$. Reference epoch $t_{\mathrm{ref}}$ is JD = 2457847.9375611655414.}
    }
    
\end{table}

\FloatBarrier 
\twocolumn

\onecolumn
\section{Stellar occultation information: Observer and events}
\label{app:2}

\begin{center}
    \begin{scriptsize}
        \begin{longtable}{lrllllr}
        \caption{ Stations details for 5044 Shestaka - 2024-10-23 }\\
        \label{tab:5044} \\
        \hline\hline
        \# & Latitude & Telescope (cm) & Method & Result & ExpTime (s) & Observer(s) \\
        Country & Longitude & Filter & Camera & Ingress (err) (s) & Deadtime (s) & \\
        Station Name & Elevation & & TimeSrc & Egress (err) (s) & & \\
        \hline
        \endfirsthead
        \caption[]{continued.}\\
        \hline
        \# & Latitude & Telescope (cm) & Method & Result & ExpTime (s) & Observer(s) \\
        Country & Longitude & Filter & Camera & Ingress (err) (s) & Deadtime (s) & \\
        Station Name & Elevation & & TimeSrc & Egress (err) (s) & & \\
        \hline
        \endhead
        \hline
        \endfoot

        1  & $+48^\circ\,48'\,22.2"$ N & T35.0 & IMG & Positive & 0.10 & V. Lapeyrère \\
        France & $+02^\circ\,13'\,47.6"$ W & IR-cut & ZWO ASI 585MC & 19:00:23.418 (0.01) & 0.02 & M. Montargès \\
        ODMP2 & 163.53 m & & ComNTP & 19:00:24.212 (0.01) & & J. Raffard et al. \\
        
        2  & $+50^\circ\,02'\,47.0"$ N & T25.0 & IMG & Positive & 0.05 & S. Quinet \\
        France & $+03^\circ\,28'\,42.4"$ W & No filter & Apollo M Max & 19:00:36.465 (0.01) & 3.0 &  \\
        Calvaire Busigny & 25.00 m & & CamGPS & 19:00:37.085 (0.01) & &  \\
        
        3  & $+50^\circ\,48'\,56.8"$ N & T20.0 & VID & Positive & 0.04 & O. Schreurs \\
        Belgium & $+04^\circ\,21'\,44.5"$ W & None & Watec 910 HX /RC & 19:00:44.756 (0.02) & ... & M. Lecossois \\
        Uccle av. Moliere & 105.80 m & & GPSBoxSprite2-VTI & 19:00:45.469 (0.02) & &  \\
        
        4  & $+45^\circ\,45'\,11.8"$ N & T20.0 & IMG & Positive & 0.05 & F. Denjean \\
        France & $-00^\circ\,55'\,57.6"$ W & NoFilter & ASI174MM & 18:59:48.192 (0.01) & ... & T. Salomon \\
        FD Spot & 12.65 m & & Other & 18:59:49.020 (0.01) & & J. Souchu \\
        
        5  & $+48^\circ\,28'\,53.2"$ N & T20.0 & IMG & Positive & 0.05 & P. L. Phan \\
        France & $+01^\circ\,56'\,46.8"$ W & Empty & Basler acA1920-40um & 19:00:20.116 (0.02) & 0.00027 &  \\
        Corbreuse & 156.00 m & & TimeBox & 19:00:20.942 (0.01) & &  \\
        
        6  & $+48^\circ\,15'\,33.2"$ N & T15.0 & IMG & Positive & 0.04 & G. Langin \\
        France & $+01^\circ\,45'\,18.0"$ W & Clear & Basler aca1920-40um ROADIES & 19:00:18.148 (0.02) & ... &  \\
        Nomad Setup & 126.70 m & & TimeBox & 19:00:18.620 (0.01) & &  \\
        
        7  & $+46^\circ\,59'\,22.6"$ N & T20.0 & VID & Positive & 0.04 & G. Rousseau \\
        France & $+00^\circ\,20'\,16.9"$ W & None & None & 19:00:02.871 (0.01) & None &  \\
        Braye-Sous-Faye & 60.00 m & & None & 19:00:03.581 (0.01) & &  \\
        
        8  & $+46^\circ\,55'\,30.6"$ N & T50.0 & IMG & Negative & 0.02 & P. Le Cam \\
        France & $+00^\circ\,47'\,17.9"$ W & No Filter & DVTI+CAM 430 &  & None &  \\
        Pressigny & 58.00 m & & CamGPS &  & &  \\
        
        9  & $+47^\circ\,13'\,23.9"$ N & T40.3 & IMG & Negative & 0.05 & P. J. Mercier \\
        France & $+00^\circ\,49'\,59.1"$ W & No filter & ZWO 294MC &  & None &  \\
        SAT37 & 91.00 m & & TimeBox &  & &  \\
        
        10  & $+48^\circ\,53'\,56.0"$ N & T30.0 & IMG & Negative & 0.04 & A. Leroy \\
        France & $+02^\circ\,42'\,20.0"$ W & no filter & QHY174M GPS &  & None &  \\
        AL station& 110.00 m & & CamGPS &  & &  \\
        
        11  & $+47^\circ\,28'\,51.4"$ N & T28.0 & IMG & Negative & 0.06 & J. L. Dumont\\
        France & $+00^\circ\,36'\,55.9"$ W & No filter & ZWO 183MM &  & None &  \\
        Nomad JLD & 51.00 m & & TimeBox &  & &  \\
        
        12  & $+48^\circ\,12'\,36.0"$ N & T26.0 & IMG & Negative & 0.03 & J. L. Dauvergne \\
        France & $+01^\circ\,44'\,32.1"$ W & empty & Player One Saturn &  & None &  \\
        Viabon & 129.00 m & & TimeBox &  & &  \\
        
        13  & $+39^\circ\,31'\,21.7"$ N & T25.6 & VID & Negative & 0.04 & R. Gonçalves \\
        Portugal& $-08^\circ\,23'\,01.7"$ W & Clear & Watec910HX-RC &  & None &  \\
        Linhaceira & 90.00 m & & Other &  & &  \\
        
        14  & $+50^\circ\,01'\,08.5"$ N & T25.0 & IMG & Negative & 0.04 & J. P. Masini \\
        France & $+03^\circ\,24'\,45.9"$ W & No Filter & VTECH 120N+ &  & None &  \\
        JPM station & 25.00 m & & IOTA-VTI &  & &  \\
        
        15  & $+50^\circ\,13'\,19.3"$ N & T20.3 & IMG & Negative & 0.03 & M. Saillenfest \\
        France & $+03^\circ\,25'\,27.9"$ W & No filter & QHY174M-GPS &  & None & A. Vienne \\
        NNVEC & 75.10 m & & CamGPS &  & &  \\
        
        16  & $+48^\circ\,28'\,12.4"$ N & T20.3 & IMG & Negative & 0.07 & T. Legault \\
        France & $+01^\circ\,49'\,32.2"$ W & None & ASI1600MM &  & None &  \\
        Orsonville & 153.18 m & & TimeBox &  & &  \\
        
        17  & $+48^\circ\,46'\,52.6"$ N & T20.0 & IMG & Negative & 0.05 & J. Desmars \\
        France & $+01^\circ\,52'\,19.0"$ W & empty & QHY174M GPS &  & None & Z. Liu \\
        Bazoches-sur-Guyonne & 114.00 m & & CamGPS &  & & D. Hestroffer \\
        
        18  & $+47^\circ\,38'\,36.0"$ N & T20.0 & IMG & Negative & 0.05 & L. Rousselot \\
        France & $+01^\circ\,10'\,50.4"$ W & None & ZWO432 &  & None &  \\
        Landes le gaulois & 113.00 m & & ComGPS &  & &  \\
        
        19  & $+46^\circ\,11'\,31.7"$ N & T20.0 & IMG & Negative & 0.05 & B. Lott \\
        France & $-00^\circ\,52'\,55.0"$ W & Clear & ZWO 1600 MM &  & None &  \\
        Anais & 17.60 m & & TimeBox &  & &  \\
        
        20  & $+47^\circ\,20'\,40.2"$ N & T20.0 & IMG & Negative & 0.04 & A. Stachowicz \\
        France & $+00^\circ\,55'\,14.2"$ W & No filter & ASI 533 MM PRO &  & None &  \\
        Nomad AS & 51.24 m & & TimeBox &  & &  \\
        
        21  & $+48^\circ\,23'\,33.6"$ N & T12.7 & IMG & Negative & 0.12 & R. Lallemand \\
        France & $+01^\circ\,42'\,04.9"$ W & Empty & QHY174M &  & None & R. Dahoumane \\
        Ablis station & 146.00 m & & CamGPS &  & & A. Ashimbekova \\
        
        22  & $+38^\circ\,57'\,15.4"$ N & T20.0 & VID & Negative & 0.04 & R. Marques \\
        Portugal & $-08^\circ\,04'\,32.4"$ W & None & WATEC 902H2 ULTIMATE (CCIR) &  & None &  \\
        Alentejo & 111.00 m & & GPS VTI &  & &  \\
        
        23  & $+50^\circ\,14'\,55.2"$ N & T27.9 & IMG & Negative & 0.03 & S. Renner \\
        France & $+03^\circ\,24'\,50.3"$ W & No filter & QHY174M-GPS &  & None &  \\
        Haspres & 58.80 m & & CamGPS &  & &  \\
        
        24  & $+49^\circ\,17'\,48.2"$ N & T25.4 & IMG & Negative & 0.04 & E. Vauthrin \\
        France & $+02^\circ\,26'\,35.9"$ W & Empty & Watec 120N+ &  & None &  \\
        Laigneville & 50.41 m & & CamGPS &  & &  \\
        
        25  & $+48^\circ\,29'\,42.5"$ N & T20.0 & IMG & Negative & 0.04 & F. Vachier \\
        France & $+02^\circ\,05'\,02.1"$ W & Empty & QHY174M-GPS &  & None &  \\
        Le Rotoir & 150.00 m & & CamGPS &  & &  \\
        
        26  & $+50^\circ\,05'\,53.4"$ N & T11.4 & IMG & Negative & 0.10 & S. Kindt \\
        France & $+03^\circ\,10'\,48.2"$ W & No filter & Sony IMX224 &  & None & P. Lemoine \\
        Masniere & 102.27 m & & ComNTP &  & &  \\
        
        27  & $+48^\circ\,44'\,04.4"$ N & T25.0 & IMG & Technical failure & 0.04 & J. Vaubaillon \\
        France & $+02^\circ\,34'\,04.7"$ W & None & Basler acA1920 &  & None &  \\
        JV station & 95.36 m & & ComGPS &  & &  \\
        
        28  & $+52^\circ\,32'\,02.5"$ N & T35.0 & VID & Overcast & ... & H. d. Groot \\
        Netherlands & $+06^\circ\,26'\,32.6"$ W & None & Watec-120N &  & None &  \\
        Ommen & 7.00 m & & Other &  & &  \\
        
        29  & $+49^\circ\,13'\,49.6"$ N & T11.4 & IMG & Overcast & 0.05 & P. Barroy \\
        France & $+02^\circ\,35'\,58.8"$ W & Empty & IMX178 &  & None &  \\
        VC PontPoint& 87.00 m & & CamGPS &  & &  \\
        
        30  & $+38^\circ\,29'\,34.7"$ N & T10.2 & VID & Technical failure & ... & R. Lourenço \\
        Portugal & $-09^\circ\,03'\,44.9"$ W & None & Watec 902 &  & None &  \\
        RL station& 109.00 m & & GPS &  & &  \\
        
        \end{longtable}
    \end{scriptsize}
\end{center}

\begin{center}
\small
    \begin{scriptsize}
        \begin{longtable}{lrllllr}
        \caption{ Stations details for 35420 1998 AG6 - 2024-07-17 }\\
        \label{ tab:35420 } \\
        \hline\hline
        \# & Latitude & Telescope (cm) & Method & Result & ExpTime (s) & Observer(s) \\
        Country & Longitude & Filter & Camera & Ingress (err) (s) & Deadtime (s) & \\
        Station Name & Elevation & & TimeSrc & Egress (err) (s) & & \\
        \hline
        \endfirsthead
        \caption[]{continued.}\\
        \hline
        \# & Latitude & Telescope (cm) & Method & Result & ExpTime (s) & Observer(s) \\
        Country & Longitude & Filter & Camera & Ingress (err) (s) & Deadtime (s) & \\
        Station Name & Elevation & & TimeSrc & Egress (err) (s) & & \\
        \hline
        \endhead
        \hline
        \endfoot

        1  & $+51^\circ\,50'\,11.6"$ N & T35.0 & IMG & Positive drop 1 & 0.04 & D. B{\l}a\.{z}ewicz \\
        Poland & $+15^\circ\,43'\,51.1"$ W & Empty & DVTI+CAM IMX430 & 01:37:24.820 (0.07) & ... &  \\
        Oty\'n& 59.00 m & & CamGPS & 01:37:25.086 (0.06) & &  \\
        
          &  &  &  & Positive drop 2 &  &  \\
         &  &  &  & 01:37:25.248 (0.06) &  &  \\
        &  & &  & 01:37:25.318 (0.07) & &  \\
        
        2  & $+47^\circ\,06'\,01.9"$ N & T20.3 & IMG & Positive & 0.08 & R. Lallemand \\
        France & $+00^\circ\,59'\,12.4"$ W & Empty & QHY174M & 01:38:11.454 (0.08) & 0.02 & J. Desmars \\
        Loches station & 129.00 m & & CamGPS & 01:38:12.015 (0.13) & &  \\
        
        3  & $+46^\circ\,50'\,40.5"$ N & T20.3 & IMG & Positive drop 1 & 0.08 & M. Saillenfest \\
        France & $+00^\circ\,12'\,52.6"$ W & Empty & QHY174M-Melaine & 01:38:14.625 (0.05) & 0.02 & Y.-N. Lee \\
        MSO & 103.00 m & & CamGPS & 01:38:14.833 (0.05) & &  \\
        
          &  &  &  & Positive drop 2 &  &  \\
         &  &  &  & 01:38:14.998 (0.06) &  &  \\
        &  & &  & 01:38:15.051 (0.11) & &  \\
        
        4  & $+39^\circ\,31'\,21.7"$ N & T25.6 & VID & Negative & 0.08 & R. Gonçalves \\
        Portugal & $-08^\circ\,23'\,01.7"$ W & Clear & Watec910HX-RC &  & 0.02 &  \\
        Linhaceira & 90.00 m & & Other &  & &  \\
        
        5  & $+47^\circ\,19'\,58.3"$ N & T20.0 & IMG & Negative & 0.05 & L. Rousselot \\
        France & $+01^\circ\,30'\,12.4"$ W & none & ZWO432 &  & 0.02 &  \\
        Chémery & 104.20 m & & ComGPS &  & &  \\
        
        6  & $+52^\circ\,16'\,37.7"$ N & T40.0 & IMG & Negative & 0.10 & A. Marciniak \\
        Poland & $+17^\circ\,04'\,23.9"$ W & Empty & Andor Zyla 5.5 sCMOS &  & 0.02 & B. Zi{\'o}{\l}kowski \\
        Borowiec & 83.00 m & & CamGPS &  & &  \\
        
        7  & $+47^\circ\,51'\,58.2"$ N & T26.0 & IMG & Overcast & 0.10 & J. L. Dauvergne \\
        France & $+03^\circ\,15'\,06.4"$ W & empty & Player One Saturn &  & 0.02 &  \\
        Touchard & 216.00 m & & TimeBox &  & &  \\

        \end{longtable}
    \end{scriptsize}
    
\end{center}

\begin{center}
\small
    \begin{scriptsize}
        \begin{longtable}{lrllllr}
        \caption{ Stations details for 206 Hersilia - 2026-01-12 }\\
        \label{ tab:206 } \\
        \hline\hline
        \# & Latitude & Telescope (cm) & Method & Result & ExpTime (s) & Observer(s) \\
        Country & Longitude & Filter & Camera & Ingress (err) (s) & Deadtime (s) & \\
        Station Name & Elevation & & TimeSrc & Egress (err) (s) & & \\
        \hline
        \endfirsthead
        \caption[]{continued.}\\
        \hline
        \# & Latitude & Telescope (cm) & Method & Result & ExpTime (s) & Observer(s) \\
        Country & Longitude & Filter & Camera & Ingress (err) (s) & Deadtime (s) & \\
        Station Name & Elevation & & TimeSrc & Egress (err) (s) & & \\
        \hline
        \endhead
        \hline
        \endfoot

        1  & $+37^\circ\,29'\,11.0"$ N & T25.0 & IMG & Positive drop 1 & 0.10 & D. Smith \\
        Spain & $-03^\circ\,02'\,17.0"$ W & None & ASI462MM & 20:37:59.570 (0.15) & ... &  \\
        Los Coloraos, Gorafe, Granada & 990.00 m & & ComNTP & 20:38:04.778 (0.15) & &  \\
        
          &  &  &  & Positive drop 2 &  &  \\
         &  &  &  & 20:38:05.749 (0.12) &  &  \\
        &  & &  & 20:38:07.856 (0.16) & &  \\
        
        2  & $+37^\circ\,06'\,41.0"$ N & T30.0 & IMG & Negative & 0.10 & D. Smith \\
        Spain & $-02^\circ\,32'\,14.0"$ W & None & ATR585M-GPS &  & ... &  \\
        Gérgal, Almería & 705.00 m & & CamGPS &  & &  \\

        \end{longtable}
    \end{scriptsize}
\end{center}

\begin{center}
\small
    \begin{scriptsize}
        \begin{longtable}{l r l l l l r}
        \caption{ Stations details for the occultation by 1127 Mimi}\\
         \label{tab:1127} \\
        \hline\hline
        \# & Latitude & Telescope (cm) & Method & Result & ExpTime (s) & Observer(s) \\
        Country & Longitude & Filter & Camera & Ingress (err) (s) & Deadtime (s) & \\
        Station Name & Elevation & & TimeSrc & Egress (err) (s) & & \\
        \hline
        \multicolumn{7}{c}{2025-02-16} \\
        \hline
        1  & $+43^\circ\,43'\,17.6"$ N & T40.0 & IMG & Positive & 0.03 & M. Conjat \\
        France & $+07^\circ\,18'\,01.4"$ W & clear & QHY 174 GPS & 01:37:36.700 (0.02) & ... &  \\
        Nice & 390.99 m & & CamGPS & 01:37:43.707 (0.02) & &  \\
        
        2  & $+37^\circ\,29'\,11.0"$ N & T28.0 & IMG & Positive & 0.10 & D. Smith \\
        Spain & $-03^\circ\,02'\,17.0"$ W & None & ASI174MM-Cool & 01:35:19.989 (0.05) & ... &  \\
        Los Coloraos, Gorafe & 990.00 m & & ComNTP & 01:35:26.750 (0.05) & &  \\
        
        3  & $+43^\circ\,09'\,20.0"$ N & T20.3 & IMG & Positive & 0.20 & P. Le Guen \\
        France & $+05^\circ\,53'\,45.9"$ W & L & ZWO ASI290MM Mini & 01:37:22.669 (0.07) & 0.05 &  \\
        Toulon-Baou & 190.45 m & & ComNTP & 01:37:29.830 (0.06) & &  \\
        
        4  & $+37^\circ\,12'\,31.0"$ N & T35.5 & IMG & Negative & 0.10 & W. Beisker \\
        Portugal & $-07^\circ\,36'\,52.0"$ W & None & DVTIcam430 &  & 0.05 &  \\
        Algarve & 148.00 m & & CamGPS &  & &  \\
        
        5  & $+50^\circ\,05'\,25.9"$ N & T40.0 & VID & Overcast & ... & R. Boninsegna \\
        Belgium & $+04^\circ\,34'\,56.0"$ W & None & Watec 910HX &  & 0.05 &  \\
        Dourbes & 192.00 m & & IOTA-VTI &  & &  \\
        
        6  & $+37^\circ\,06'\,41.0"$ N & T30.0 & IMG & Overcast & ... & D. Smith \\
        Spain & $-02^\circ\,32'\,14.0"$ W & Baader UVIR cut 2" & ASI533MM &  & 0.05 &  \\
        Gérgal, Almería & 705.00 m & & ComNTP &  & &  \\
        
        7  & $+36^\circ\,52'\,50.8"$ N & T20.0 & VID & Overcast & 0.02 & F. Casarramona \\
        Spain & $-02^\circ\,00'\,54.7"$ W & Empty & Watec 910HX/RC &  & 0.05 &  \\
        Las Negras & 62.19 m & & Kiwi VTI &  & &  \\
        
        \hline
        \multicolumn{7}{c}{2025-02-26} \\
        \hline
        1  & $+41^\circ\,37'\,16.4"$ N & T27.8 & IMG & Positive & 0.10 & O. Canales \\
        Spain & $-00^\circ\,57'\,08.6"$ W & Clear & QHY174-GPS & 19:33:12.762 (0.11) & ... &  \\
        Observatorio Arcosur & 268.75 m & & CamGPS & 19:33:19.039 (0.11) & &  \\
        
        2  & $+48^\circ\,49'\,18.0"$ N & T20.3 & IMG & Positive & 0.50 & G. Langin \\
        France & $+02^\circ\,20'\,06.4"$ W & Empty & Basler aca1920-40um & 19:35:23.093 (0.70) & ... & P. Henarejos \\
        Montsouris AFA & 77.10 m & & TimeBox & 19:35:27.442 (0.60) & &  \\
        
        3  & $+44^\circ\,35'\,02.4"$ N & T20.3 & IMG & Positive & 0.12 & J. Souchu \\
        France & $-00^\circ\,14'\,51.7"$ W & Sans filtre & Zwo Asi 178mm & 19:34:06.524 (0.4) & 0.033 & C. Bourdens \\
        Moulin de Cussol & 30.00 m & & Other & 19:34:09.286 (0.21) & & D. Bourdens \\
        
        4  & $+45^\circ\,31'\,01.8"$ N & T50.8 & IMG & Negative & 0.10 & E. Barbotin \\
        France & $-00^\circ\,00'\,26.4"$ W & Clear & ZWO6200 &  & 0.033 &  \\
        ODLGVE & 119.00 m & & ComNTP &  & &  \\
        
        5  & $+46^\circ\,13'\,53.1"$ N & T42.0 & IMG & Negative & 0.10 & S. Sposetti \\
        Switzerland & $+09^\circ\,01'\,26.6"$ W & None & DVTI &  & 0.033 &  \\
        Gnosca & 258.88 m & & CamGPS &  & &  \\
        
        6  & $+41^\circ\,29'\,37.0"$ N & T40.0 & VID & Negative & 0.08 & C. Schnabel \\
        Spain & $+01^\circ\,52'\,21.0"$ W & Empty & WATEC-910HX/RC &  & 0.033 &  \\
        Sant Esteve Sesrovires & 180.00 m & & IOTA-VTI &  & &  \\
        
        7  & $+42^\circ\,42'\,42.2"$ N & T31.0 & VID & Negative & 0.16 & P. Martorell \\
        Spain & $-01^\circ\,51'\,54.4"$ W & None & Watec-120N+ &  & 0.033 &  \\
        OADG & 594.03 m & & CamGPS &  & &  \\
        
        8  & $+51^\circ\,10'\,20.3"$ N & T35.6 & IMG & Overcast & ... & J. Bourgeois \\
        Belgium & $+03^\circ\,26'\,10.2"$ W & clear & DVTI-CAM 430 &  & 0.033 &  \\
        Kleit Astronomical Station & 24.00 m & & CamGPS &  & &  \\
        
        9  & $+49^\circ\,17'\,48.2"$ N & T25.4 & IMG & Overcast & ... & E. Vauthrin \\
        France & $+02^\circ\,26'\,35.9"$ W & Empty & Watec 120N+ &  & 0.033 &  \\
        Laigneville & 50.41 m & & CamGPS &  & &  \\
        
        10  & $+51^\circ\,20'\,08.9"$ N & T20.0 & IMG & Overcast & ... & O. Schreurs \\
        Belgium & $+03^\circ\,16'\,48.6"$ W & Clear & - &  & 0.033 &  \\
        Knokke Station & 0.00 m & & - &  & &  \\
        
        11  & $+37^\circ\,06'\,41.0"$ N & T30.0 & IMG & Technical failure & ... & D. Smith \\
        Spain & $-02^\circ\,32'\,14.0"$ W & Baader UVIR cut 2" & ASI533MM &  & 0.033 &  \\
        Gérgal, Almería & 705.00 m & & ComNTP &  & &  \\
        
        12  & $+37^\circ\,29'\,11.0"$ N & T28.0 & IMG & Technical failure & ... & D. Smith \\
        Spain & $-03^\circ\,02'\,17.0"$ W & None & ASI174MM-Cool &  & 0.033 &  \\
        Los Coloraos, Gorafe & 990.00 m & & ComNTP &  & &  \\
        
        13  & $+42^\circ\,13'\,17.4"$ N & T20.0 & IMG & Technical failure & ... & J. Prat \\
        Spain & $-01^\circ\,41'\,19.9"$ W & None & QHY-174M-GPS &  & 0.033 &  \\
        Anko & 284.08 m & & CamGPS &  & &  \\
        \hline
        \end{longtable}
    \end{scriptsize}
\end{center}

\begin{center}
\small
    \begin{scriptsize}
        \begin{longtable}{lrllllr}
        \caption{ stations details for 36882 2000 SW155 - 2025-08-29 }\\
        \label{tab:36882} \\
        \hline\hline
        \# & Latitude & Telescope (cm) & Method & Result & ExpTime (s) & Observer(s) \\
        Country & Longitude & Filter & Camera & Ingress (err) (s) & Deadtime (s) & \\
        Station Name & Elevation & & TimeSrc & Egress (err) (s) & & \\
        \hline
        \endfirsthead
        \caption[]{continued.}\\
        \hline
        \# & Latitude & Telescope (cm) & Method & Result & ExpTime (s) & Observer(s) \\
        Country & Longitude & Filter & Camera & Ingress (err) (s) & Deadtime (s) & \\
        Station Name & Elevation & & TimeSrc & Egress (err) (s) & & \\
        \hline
        \endhead
        \hline
        \endfoot
        
        1  & $+36^\circ\,52'\,50.8"$ N & T20.0 & VID & Positive drop 1 & 0.08 & F. Casarramona \\
        Spain & $-02^\circ\,00'\,54.7"$ W & Empty & Watec 910HX/RC & 22:52:55.216 (0.01) & ... &  \\
        Las Negras & 62.19 m & & IOTA-VTI & 22:52:55.336 (0.03) & &  \\
        
          &  &  &  & Positive drop 2 &  &  \\
         &  &  &  & 22:52:55.746 (0.02) &  &  \\
         &  & &  & 22:52:55.834 (0.02) & &  \\
        
        2  & $+52^\circ\,22'\,57.3"$ N & T20.3 & IMG & Positive & 0.03 & D. Antuszewicz \\
        Poland & $+20^\circ\,54'\,09.0"$ W & Empty & ZWO ASI715MC & 22:50:11.415 (0.06) & 0.001 &  \\
        Home Legionowo & 80.08 m & & ComNTP & 22:50:12.334 (0.05) & &  \\
        
        3  & $+43^\circ\,43'\,32.9"$ N & T40.0 & IMG & Negative & 0.02 & M. Conjat \\
        France & $+07^\circ\,17'\,59.4"$ W & clear & QHY 174 GPS &  & ... &  \\
        Observatoire de Nice & 320.00 m & & CamGPS &  & &  \\
        
        4  & $+43^\circ\,09'\,20.0"$ N & T10.6 & IMG & Negative & 0.03 & P. Le Guen \\
        France & $+05^\circ\,53'\,45.9"$ W & L & ZWO ASI6200MM-PRO &  & ... &  \\
        Toulon-Baou & 190.45 m & & ComNTP &  & &  \\
        
        5  & $+53^\circ\,06'\,32.2"$ N & T60 & IMG & Negative & 0.03 & W. Burzynski \\
        Poland & $+23^\circ\,09'\,21.2"$ W & L & DVTI+CAM &  & ... &  \\
        Bialystok & 176.00 m & & GPS &  & &  \\
        
        6  & $+43^\circ\,09'\,01.5"$ N & T20 & CCD & Negative & 0.04 & J.-F. Coliac \\
        France & $+05^\circ\,50'\,25.8"$ W & None & Watec 910 HX &  & ... & F. Gourdon \\
        OAGC & 200.00 m & & VTI &  & &  \\
        
        7  & $+46^\circ\,59'\,33.0"$ N & T35.6 & IMG & Overcast & ... & C. Ziolek \\
        Switzerland & $+09^\circ\,38'\,24.4"$ W & none & DVTI+CAM 430 (4x4) &  & ... &  \\
        Sternwarte Seewis Dorf & 968.07 m & & CamGPS &  & &  \\
        
        \end{longtable}
        
    \end{scriptsize}
\end{center}

\FloatBarrier 
\twocolumn

\onecolumn
\section{Positive light curves}
\label{app:3}

\begin{figure}[ht!]
\centering
\includegraphics[width=\textwidth]{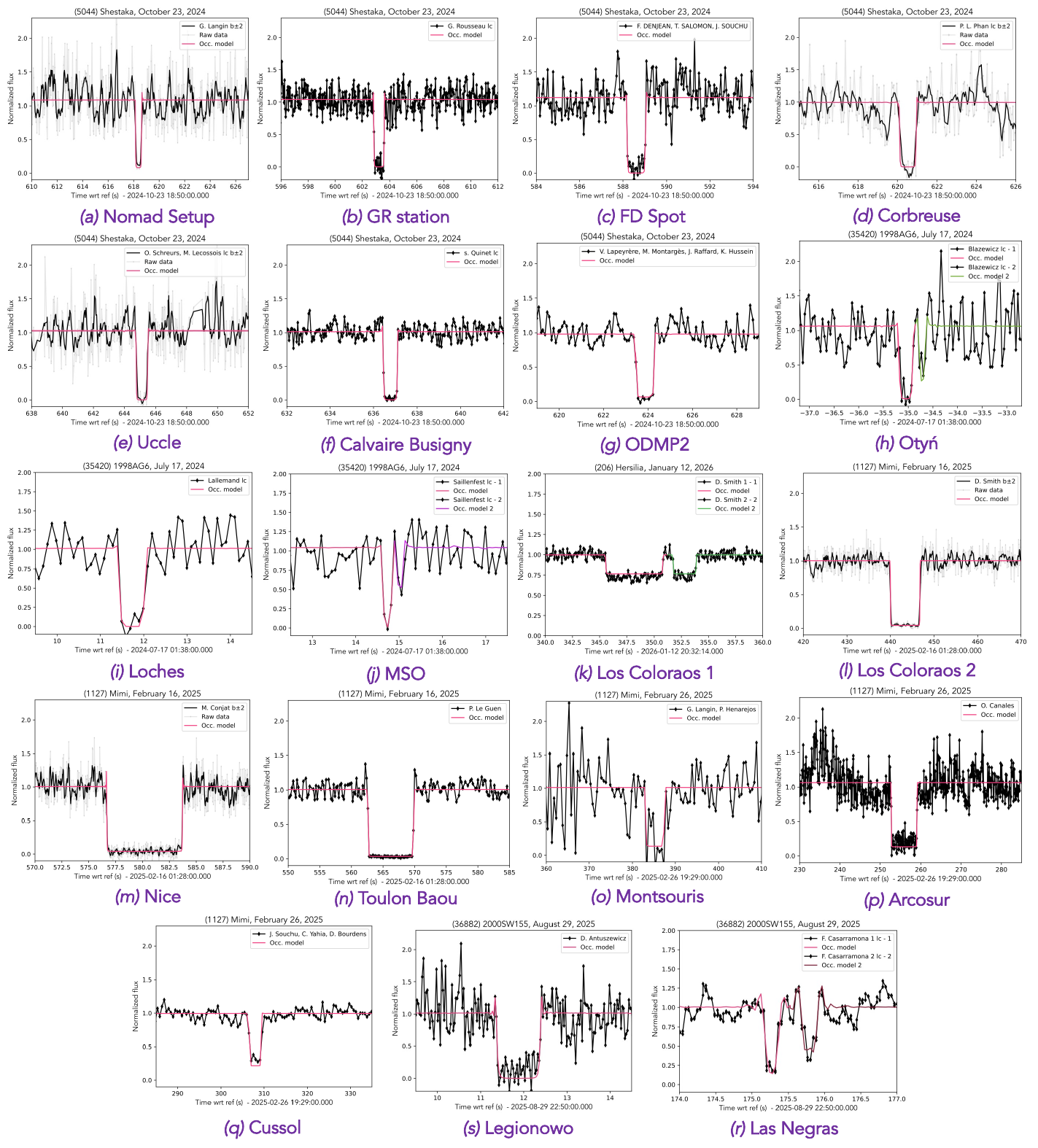}
  \caption{Normalised positive light curves from stellar occultation events presented in \ref{sec:results}. The associated event is presented in the title of the light curve. Raw data are represented in black line and fitted occultation model is represented in pink. The date and observational Station are indicated in the label. For certain light curve, the mention "$n\pm$" indicates binning for better visualisation. After analysis, the first drop measure by (o) Montsouris is mixed noise.}
     \label{figA4}
\end{figure}

\FloatBarrier 
\twocolumn

\onecolumn

\section{Occultation fit results}
\label{app:5}

\begin{figure}[ht!]
\centering
\includegraphics[width = \hsize]{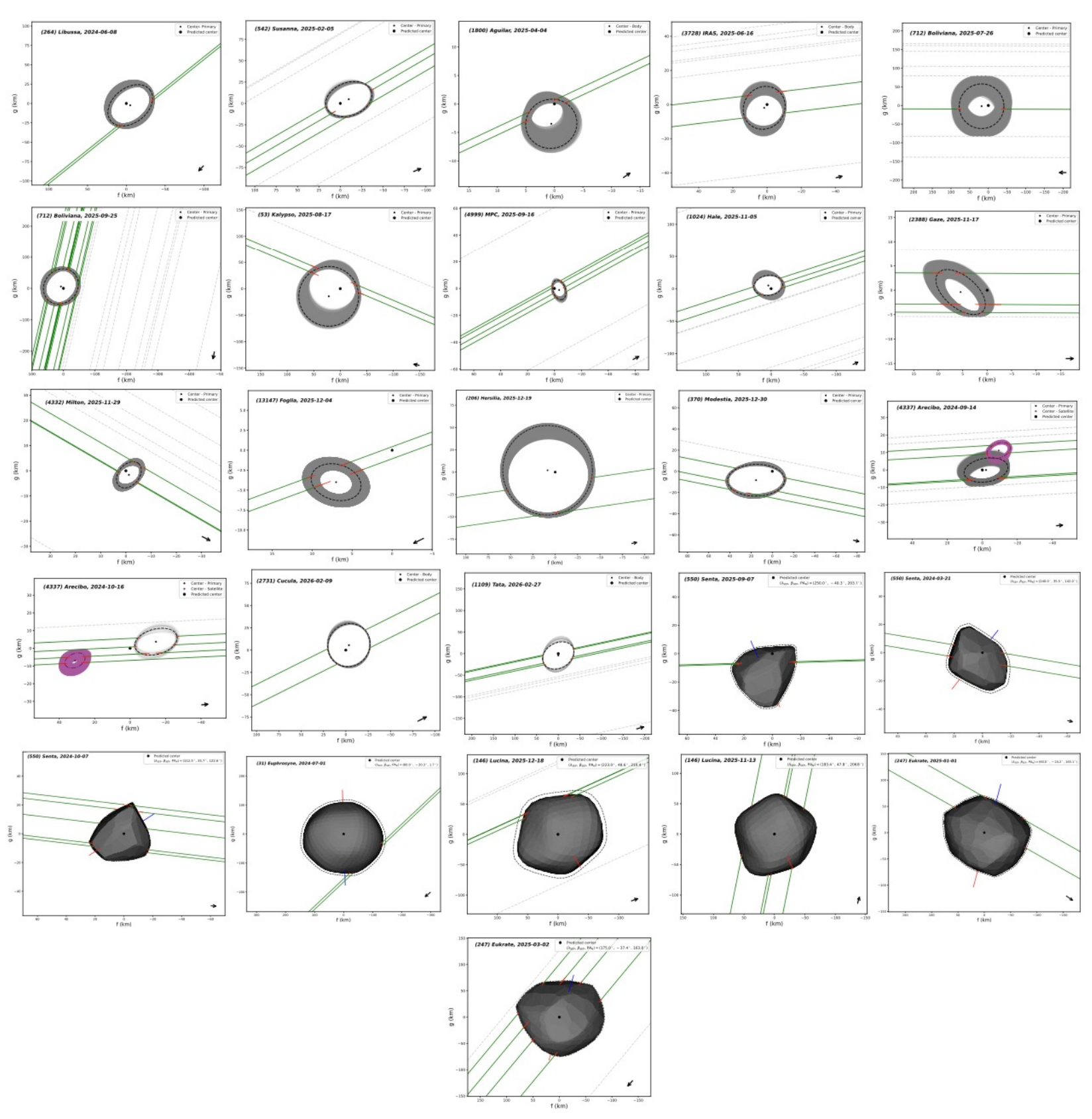}

  \caption{Results of elliptical or shape fit for stellar occultation events listed on Table~\ref{tableCampaign}. The name of the object and the occultation date are displayed in the upper left side of the plots. For elliptical fits, solutions within the uncertainty are given in grey ellipses. For shape fits, uncertainties are given as a dotted-line limb around the shape. The satellites position and uncertainties are displayed in purple. Captions are the same as for Figure~\ref{fig2} left panel.}
     \label{figA51}
\end{figure}

\FloatBarrier 
\twocolumn

\onecolumn
\section{Occultation paths}
\label{app:4}

\begin{figure}[ht!]
\centering
\includegraphics[width=0.9\textwidth]{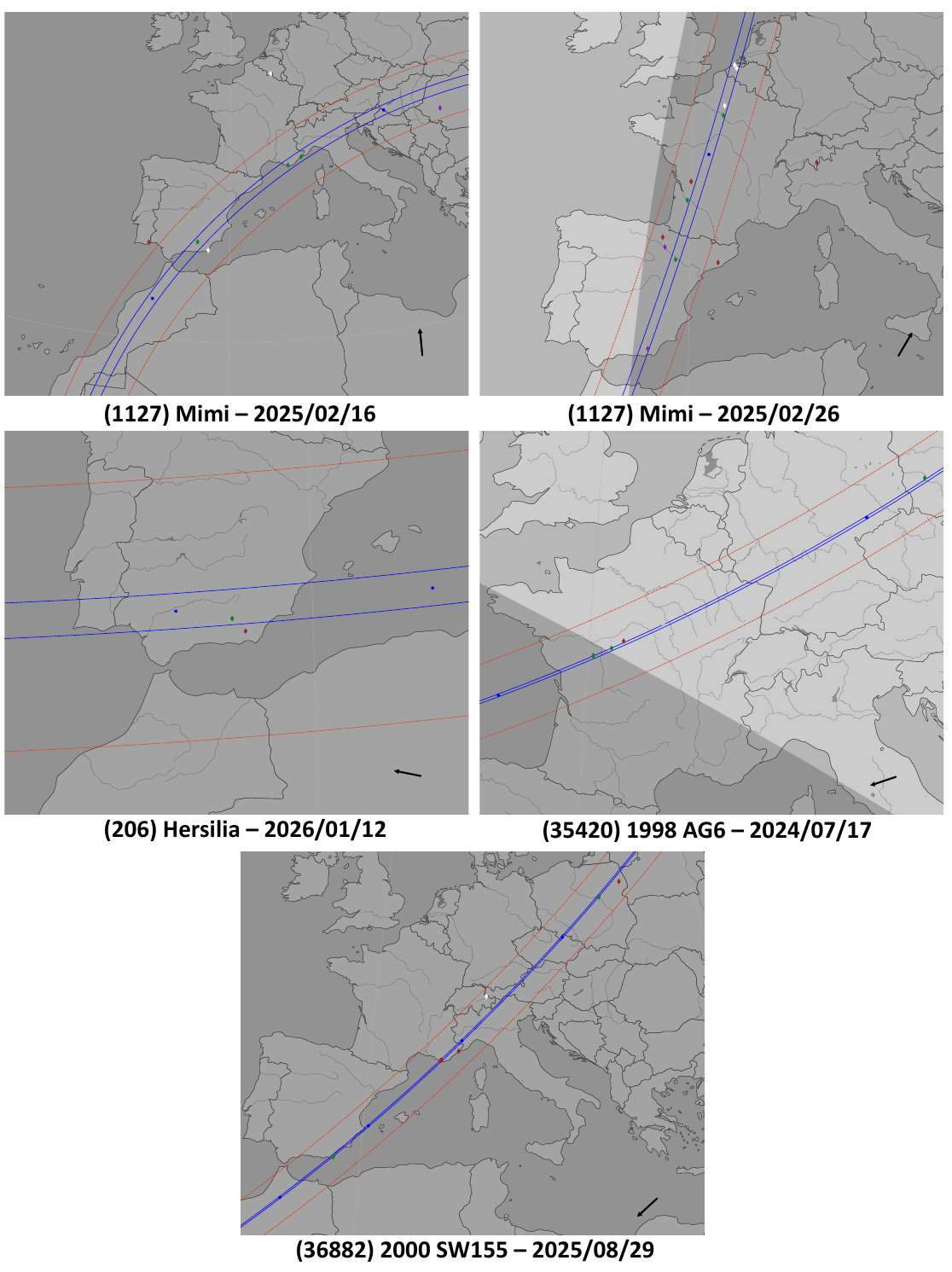}
  \caption{Post-event map after observation for each relevant stellar occultation event. Legends are the same as the post-occultation map described in Figure~\ref{fig2}. The shadow follows the direction given by the black arrow.}
     \label{figA6}
\end{figure}

\FloatBarrier 
\twocolumn

\end{appendix}

\end{document}